\documentclass[TOTEM]{style/cern/cernphprep}

\usepackage[utf8]{inputenc}
\usepackage{color}
\usepackage{mathptmx}
\RequirePackage[numbers,sort&compress]{natbib}
\RequirePackage[colorlinks,citecolor=blue,urlcolor=blue,linkcolor=blue]{hyperref}
\RequirePackage{amsmath}
\RequirePackage{amssymb}
\RequirePackage{enumitem}

\DeclareMathAlphabet{\mathcal}{OMS}{cmsy}{m}{n}

\def\d{{\rm d}}
\def\un#1{\,{\rm #1}}
\def\ung#1{\quad[{\rm #1}]}
\def\unt#1{[{\rm #1}]}

\def\I{{\rm i}}
\def\T{{\rm T}}

\def\mat#1{\mathsf{#1}}
\def\etal{et al.}

\newskip\itskip \itskip2mm
\newskip\iitskip \iitskip0mm
\newskip\iiitskip \iiitskip0mm

\newdimen\itindent \itindent4mm
\newdimen\iitindent \iitindent8mm
\newdimen\iiitindent \iiitindent12mm

\bgroup
\catcode`\>=11

\gdef\>{%
	\par\vskip1mm
	\par\noindent\hbox to 2mm{\hss $\bullet$}
}

\gdef\>>{%
	\par
	\noindent\hbox to 7mm{\hss $\circ$}
}

\gdef\>>>{%
	\par
	\noindent\hbox to 12mm{\hss --}
}

\egroup

\begin{document}

\begin{titlepage}

\renewcommand{\EXPLOGO}{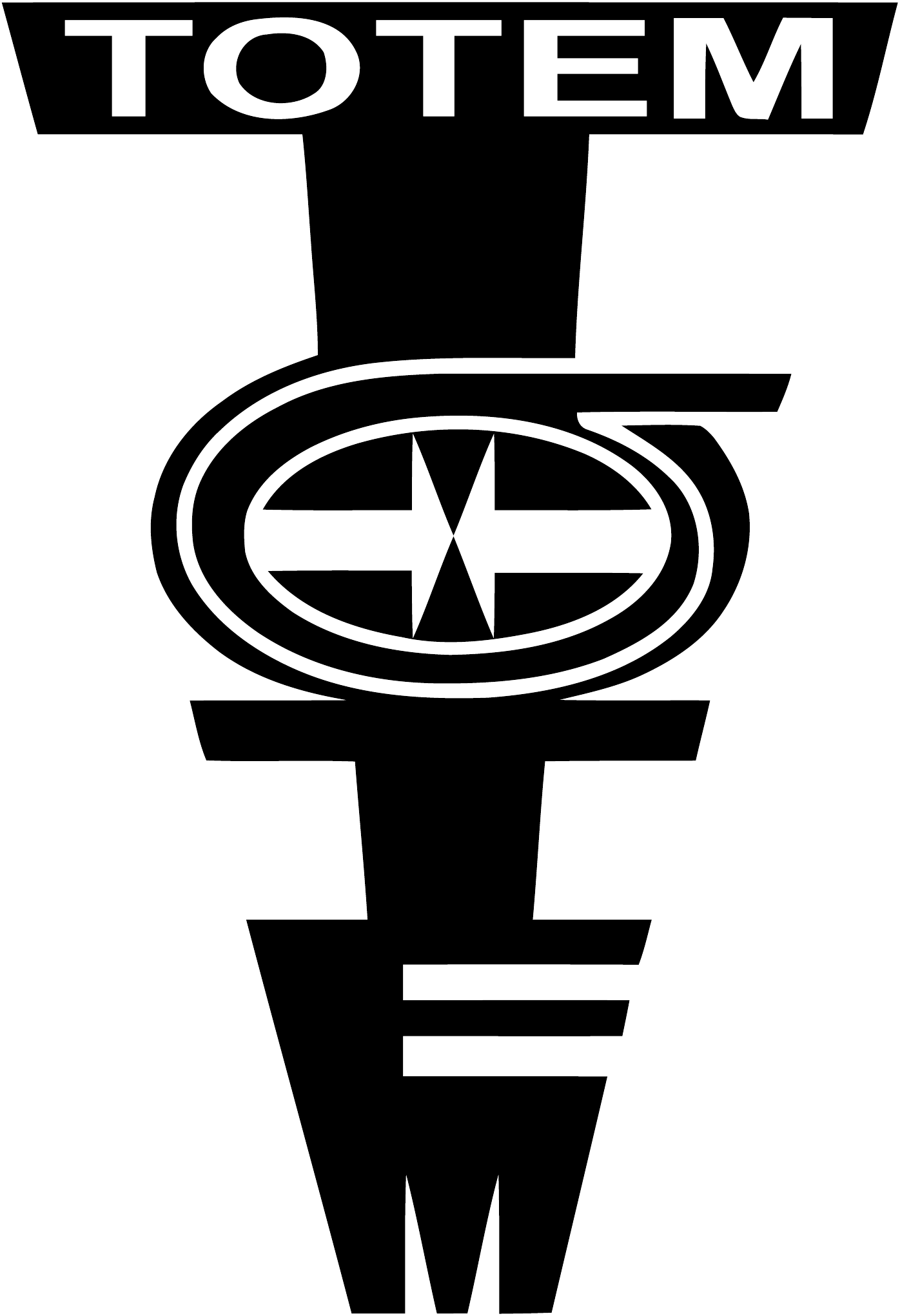}

\PHnumber{CERN-EP-2017-335-v3}
\PHdate{30 November 2018}

\EXPnumber{TOTEM-2017-002}
\EXPdate{30 November 2018}

	\title{First determination of the ${\rho}$ parameter at ${\sqrt{ s} = 13}$ TeV -- probing the existence of a colourless three-gluon bound state}

\ShortTitle{First determination of $\rho$ at $\sqrt s = 13$ TeV -- probing a colourless 3-gluon bound state}

\Collaboration{The TOTEM Collaboration}

\ShortAuthor{The TOTEM Collaboration (G.~Antchev \emph{\etal})}



\newif\ifFirstAuthor
\FirstAuthortrue

\def\AddAuthor#1#2#3#4{%
	\def\PriAf{#2}%
	\def\SecAf{#3}%
	\def\ExtAf{#4}%
	\def\empty{}%
	\ifFirstAuthor
		\FirstAuthorfalse
	\else
		,
	\fi
	\ifx\PriAf\empty
		#1\Aref{#4}%
	\else
		\ifx\SecAf\empty
			\ifx\ExtAf\empty
				#1\Iref{#2}%
			\else
				#1\IAref{#2}{#4}%
			\fi
		\else
			\ifx\ExtAf\empty
				#1\IIref{#2}{#3}%
			\else
				#1\IIAref{#2}{#3}{#4}%
				\relax
			\fi
		\fi
	\fi
}


\def\AddCorrespondingAuthor#1#2#3#4#5#6{%
	\AddAuthor{#1}{#2}{#3}{*}%
	\Anotfoot{*}{#5 E-mail address: #6.}
}


\def\AddInstitute#1#2{%
	\expandafter\write1{\string\newlabel{#1}{{#1}{}}}%
	\hbox to\hsize{\strut\hss$^{#1}$#2\hss}%
}


\def\AddExternalInstitute#1#2{%
	\Anotfoot{#1}{#2}%
}



\def\DeclareAuthors{%
	\AddAuthor{G.~Antchev}{}{}{a}%
	\AddAuthor{P.~Aspell}{9}{}{}%
	\AddAuthor{I.~Atanassov}{}{}{a}%
	\AddAuthor{V.~Avati}{7}{9}{}%
	\AddAuthor{J.~Baechler}{9}{}{}%
	\AddAuthor{C.~Baldenegro Barrera}{11}{}{}%
	\AddAuthor{V.~Berardi}{4a}{4b}{}%
	\AddAuthor{M.~Berretti}{2a}{}{}%
	\AddAuthor{E.~Bossini}{6b}{}{}%
	\AddAuthor{U.~Bottigli}{6b}{}{}%
	\AddAuthor{M.~Bozzo}{5a}{5b}{}%
	\AddAuthor{R.~Bruce}{9}{}{}%
	\AddAuthor{H.~Burkhardt}{9}{}{}%
	\AddAuthor{F.~S.~Cafagna}{4a}{}{}%
	\AddAuthor{M.~G.~Catanesi}{4a}{}{}%
	\AddAuthor{M.~Csan\'{a}d}{3a}{}{b}%
	\AddAuthor{T.~Cs\"{o}rg\H{o}}{3a}{3b}{}%
	\AddAuthor{M.~Deile}{9}{}{}%
	\AddAuthor{F.~De~Leonardis}{4c}{4a}{}%
	\AddAuthor{A.~D'Orazio}{4c}{4a}{}%
	\AddAuthor{M.~Doubek}{1c}{}{}%
	\AddAuthor{D.~Druzhkin}{9}{}{}%
	\AddAuthor{K.~Eggert}{10}{}{}%
	\AddAuthor{V.~Eremin}{}{}{e}%
	\AddAuthor{F.~Ferro}{5a}{}{}%
	\AddAuthor{A.~Fiergolski}{9}{}{}%
	\AddAuthor{F.~Garcia}{2a}{}{}%
	\AddAuthor{H.~Garcia Morales}{9}{}{h}%
	\AddAuthor{V.~Georgiev}{1a}{}{}%
	\AddAuthor{S.~Giani}{9}{}{}%
	\AddAuthor{L.~Grzanka}{7}{}{c}%
	\AddAuthor{J.~Hammerbauer}{1a}{}{}%
	\AddAuthor{J.~Heino}{2a}{}{}%
	\AddAuthor{P.~Helander}{2a}{2b}{}%
	\AddAuthor{T.~Isidori}{11}{}{}%
	\AddAuthor{V.~Ivanchenko}{8}{}{}%
	\AddAuthor{M.~Janda}{1c}{}{}%
	\AddAuthor{A.~Karev}{9}{}{}%
	\AddAuthor{J.~Ka\v{s}par}{6a}{1b}{}%
	\AddAuthor{J.~Kopal}{9}{}{}%
	\AddAuthor{V.~Kundr\'{a}t}{1b}{}{}%
	\AddAuthor{S.~Lami}{6a}{}{}%
	\AddAuthor{G.~Latino}{6b}{}{}%
	\AddAuthor{R.~Lauhakangas}{2a}{}{}%
	\AddAuthor{R.~Linhart}{1a}{}{}%
	\AddAuthor{C.~Lindsey}{11}{}{}%
	\AddAuthor{M.~V.~Lokaj\'{\i}\v{c}ek}{1b}{}{}%
	\AddAuthor{L.~Losurdo}{6b}{}{}%
	\AddAuthor{M.~Lo~Vetere}{5b}{5a}{+}%
	\AddAuthor{F.~Lucas~Rodr\'{i}guez}{9}{}{}%
	\AddAuthor{M.~Macr\'{\i}}{5a}{}{}%
	\AddAuthor{M.~Malawski}{7}{}{}%
	\AddAuthor{A.~Mereghetti}{9}{}{}%
	\AddAuthor{N.~Minafra}{11}{}{}%
	\AddAuthor{S.~Minutoli}{5a}{}{}%
	\AddAuthor{D.~Mirarchi}{9}{}{}%
	\AddAuthor{T.~Naaranoja}{2a}{2b}{}%
	\AddAuthor{F.~Nemes}{9}{3a}{}%
	\AddAuthor{H.~Niewiadomski}{10}{}{}%
	\AddAuthor{T.~Nov\'{a}k}{3b}{}{}%
	\AddAuthor{E.~Oliveri}{9}{}{}%
	\AddAuthor{F.~Oljemark}{2a}{2b}{}%
	\AddAuthor{M.~Oriunno}{}{}{f}%
	\AddAuthor{K.~\"{O}sterberg}{2a}{2b}{}%
	\AddAuthor{P.~Palazzi}{9}{}{}%
	\AddAuthor{V.~Passaro}{4c}{4a}{}%
	\AddAuthor{Z.~Peroutka}{1a}{}{}%
	\AddAuthor{J.~Proch\'{a}zka}{1b}{}{}%
	\AddAuthor{M.~Quinto}{4a}{4b}{}%
	\AddAuthor{E.~Radermacher}{9}{}{}%
	\AddAuthor{E.~Radicioni}{4a}{}{}%
	\AddAuthor{F.~Ravotti}{9}{}{}%
	\AddAuthor{S.~Redaelli}{9}{}{}%
	\AddAuthor{E.~Robutti}{5a}{}{}%
	\AddAuthor{C.~Royon}{11}{}{}%
	\AddAuthor{G.~Ruggiero}{9}{}{}%
	\AddAuthor{H.~Saarikko}{2a}{2b}{}%
	\AddAuthor{B.~Salvachua}{9}{}{}%
	\AddAuthor{A.~Scribano}{6a}{}{}%
	\AddAuthor{J.~Siroky}{1a}{}{}%
	\AddAuthor{J.~Smajek}{9}{}{}%
	\AddAuthor{W.~Snoeys}{9}{}{}%
	\AddAuthor{R.~Stefanovitch}{9}{}{}%
	\AddAuthor{J.~Sziklai}{3a}{}{}%
	\AddAuthor{C.~Taylor}{10}{}{}%
	\AddAuthor{E.~Tcherniaev}{8}{}{}%
	\AddAuthor{N.~Turini}{6b}{}{}%
	\AddAuthor{V.~Vacek}{1c}{}{}%
	\AddAuthor{G.~Valentino}{}{}{g}%
	\AddAuthor{J.~Welti}{2a}{2b}{}%
	\AddAuthor{J.~Wenninger}{9}{}{}%
	\AddAuthor{J.~Williams}{11}{}{}%
	\AddAuthor{P.~Wyszkowski}{7}{}{}%
	\AddAuthor{J.~Zich}{1a}{}{}%
	\AddAuthor{K.~Zielinski}{7}{}{}%
}


\def\DeclareInstitutes{%
	\AddInstitute{1a}{University of West Bohemia, Pilsen, Czech Republic.}
	\AddInstitute{1b}{Institute of Physics of the Academy of Sciences of the Czech Republic, Prague, Czech Republic.}
	\AddInstitute{1c}{Czech Technical University, Prague, Czech Republic.}
	\AddInstitute{2a}{Helsinki Institute of Physics, University of Helsinki, Helsinki, Finland.}
	\AddInstitute{2b}{Department of Physics, University of Helsinki, Helsinki, Finland.}
	\AddInstitute{3a}{Wigner Research Centre for Physics, RMKI, Budapest, Hungary.}
	\AddInstitute{3b}{EKU KRC, Gy\"ongy\"os, Hungary.}
	\AddInstitute{4a}{INFN Sezione di Bari, Bari, Italy.}
	\AddInstitute{4b}{Dipartimento Interateneo di Fisica di Bari, Bari, Italy.}
	\AddInstitute{4c}{Dipartimento di Ingegneria Elettrica e dell'Informazione - Politecnico di Bari, Bari, Italy.}
	\AddInstitute{5a}{INFN Sezione di Genova, Genova, Italy.}
	\AddInstitute{5b}{Universit\`{a} degli Studi di Genova, Italy.}
	\AddInstitute{6a}{INFN Sezione di Pisa, Pisa, Italy.}
	\AddInstitute{6b}{Universit\`{a} degli Studi di Siena and Gruppo Collegato INFN di Siena, Siena, Italy.}
	\AddInstitute{7}{AGH University of Science and Technology, Krakow, Poland.}
	\AddInstitute{8}{Tomsk State University, Tomsk, Russia.}
	\AddInstitute{9}{CERN, Geneva, Switzerland.}
	\AddInstitute{10}{Case Western Reserve University, Dept.~of Physics, Cleveland, OH, USA.}
	\AddInstitute{11}{The University of Kansas, Lawrence, USA.}
}

	
\def\DeclareExternalInstitutes{%
	\AddExternalInstitute{a}{INRNE-BAS, Institute for Nuclear Research and Nuclear Energy, Bulgarian Academy of Sciences, Sofia, Bulgaria.}
	\AddExternalInstitute{b}{Department of Atomic Physics, ELTE University, Budapest, Hungary.}
	\AddExternalInstitute{c}{Institute of Nuclear Physics, Polish Academy of Science, Krakow, Poland.}
	\AddExternalInstitute{d}{Warsaw University of Technology, Warsaw, Poland.}
	\AddExternalInstitute{e}{Ioffe Physical - Technical Institute of Russian Academy of Sciences, St.~Petersburg, Russian Federation.}
	\AddExternalInstitute{f}{SLAC National Accelerator Laboratory, Stanford CA, USA.}
	\AddExternalInstitute{g}{University of Malta, Msida, Malta.}
	\AddExternalInstitute{h}{Royal Holloway University of London, Egham, UK.}
	\AddExternalInstitute{+}{Deceased.}
}
\begin{Authlist}
	\DeclareAuthors
\end{Authlist}

\DeclareInstitutes
\hbox to\hsize{\strut\hss} 
\DeclareExternalInstitutes

\begin{abstract}
	The TOTEM experiment at the LHC has performed the first measurement at
	$\sqrt s = 13\un{TeV}$ of the $\rho$ parameter, the real to imaginary
	ratio of the nuclear elastic scattering amplitude at $t=0$, obtaining
	the following results: $\rho = 0.09 \pm 0.01$ and $\rho = 0.10 \pm
	0.01$, depending on different physics assumptions and mathematical
	modelling. The unprecedented precision of the $\rho$ measurement,
	combined with the TOTEM total cross-section measurements in an energy
	range larger than $10\un{TeV}$ (from $2.76$ to $13\un{TeV}$), has
	implied the exclusion of all the models classified and published by
	COMPETE. The $\rho$ results obtained by TOTEM are compatible with the
	predictions, from alternative theoretical models both in the Regge-like
	framework and in the QCD framework, of a colourless 3-gluon bound state
	exchange in the $t$-channel of the proton-proton elastic scattering. On
	the contrary, if shown that the 3-gluon bound state $t$-channel
	exchange is not of importance for the description of elastic
	scattering, the $\rho$ value determined by TOTEM would represent a
	first evidence of a slowing down of the total cross-section growth at
	higher energies.
	The very low-$|t|$ reach allowed also to determine the absolute normalisation
	using the Coulomb amplitude for the first time at the LHC and obtain a
	new total proton-proton cross-section measurement $\sigma_{\rm tot} =
	(110.3 \pm 3.5)\un{mb}$, completely independent from the previous TOTEM
	determination. Combining the two TOTEM results yields $\sigma_{\rm tot}
	= (110.5 \pm 2.4)\un{mb}$.
\end{abstract}
\end{titlepage}





\section{Introduction}
\label{sec:introduction}

The TOTEM experiment at the LHC has measured the differential elastic
proton-proton scattering cross-section as a function of the four-momentum
transfer squared, $t$, down to $|t| = 8\times10^{-4}\un{GeV^2}$ at the
centre-of-mass energy $\sqrt s = 13\un{TeV}$ using a special $\beta^* =
2.5\un{km}$ optics. This allowed to access the Coulomb-nuclear interference
(CNI) and to determine the real-to-imaginary ratio of the forward hadronic
amplitude with an unprecedented precision.

Measurements of the total proton-proton cross-section and $\rho$ have been
published in the literature from the low energy range of $\sqrt s \sim
10\un{GeV}$ up to the LHC energy of $8\un{TeV}$ \cite{pdg-2016}. Such
experimental measurements have been parametrised by a large variety of
phenomenological models in the last decades, and were analysed and classified
by the COMPETE collaboration \cite{compete}.

It is shown in the present paper that none of the above-mentioned models can
describe simultaneously the TOTEM $\rho$ measurement at $13\un{TeV}$ and the
ensemble of the total cross-section measurements by TOTEM ranging from $\sqrt s
= 2.76$ to $13\un{TeV}$
\cite{totem-7tev-tot2,totem-8tev-90m,totem-8tev-1km,totem-13tev-90m}. The
exclusion of the COMPETE published models is quantitatively demonstrated on the
basis of the p-values reported in this work. Such conventional modelling of the
low-$|t|$ nuclear elastic scattering is based on various forms of Pomeron
exchanges and related crossing-even scattering amplitudes (not changing sign
under the crossing symmetry, cf.~Section 4.5~in \cite{barone-predazzi}).

Sophisticated alternative theoretical models exist both in terms of Regge-like
or axiomatic field theories \cite{nicolescu-1992} and of QCD \cite{braun} --
they are capable of predicting or taking into account several effects confirmed
or observed at LHC energies: the existence of a sharp diffractive dip in the
proton-proton elastic t-distribution also at LHC energies
\cite{totem-7tev-first}, the deviation of the elastic differential
cross-section from a pure exponential \cite{totem-8tev-90m}, the deviation of
the elastic diffractive slope, $B$, from a linear $\log(s)$ dependence as a
function of the centre-of-mass energy \cite{totem-13tev-90m}, the variation of
the nuclear phase as a function of $t$, the large-$|t|$ power-law behaviour of
the elastic $t$-distribution with no oscillatory behaviour and the growth rate
of the total cross-section as a function of $\sqrt s$ at LHC energies
\cite{totem-13tev-90m}. These theoretical frameworks foresee the possibility of
more complex $t$-channel exchanges in the proton-proton elastic interaction,
including crossing-odd scattering amplitude contributions (changing sign under
the crossing symmetry).

The crossing-odd contributions were associated with the concept of the Odderon
(the crossing-odd counterpart of the Pomeron \cite{lipatov-1986}) invented in
the '70s \cite{nicolescu-1973,nicolescu-1975} and later confirmed as an
essential QCD prediction
\cite{bartels-1980,kwiecinski-1980,jaroszewicz-1981,ioffe-2010}. They are
quantified in QCD \cite{levin-1990,durham-2018} where they are represented (in
the most basic form) by the exchange of a colourless 3-gluon bound state in the
t-channel. Such a state would naturally have $J^{PC}=1^{--}$ quantum numbers
and is predicted by lattice QCD with a mass of about $3$ to $4\un{GeV}$ (also
referred to as vector glueball).

Experimental searches for such state have used various channels. In central
production the 3-gluon state emitted by one proton may fuse with a Pomeron or
photon emitted from the other proton and create a detectable meson system
\cite{hera-odderon-2002}. However, such processes are dominated by Pomeron(s)
and Pomeron-photon exchanges, making the observation of a 3-gluon state
difficult. In elastic scattering at low energy \cite{breakstone-85}, the
observation of 3-gluon bound state is complicated by the presence of secondary
Regge trajectories influencing the potential observation of differences between
the proton-proton and proton-antiproton scattering. At high energy
(gluonic-dominated interactions) \cite{yellow-report}, one could investigate
for both proton-proton and proton-antiproton scattering the diffractive dip,
where the imaginary part of the Pomeron amplitude vanishes; however there are
no measurements nor facilities allowing a comparison at the same fixed $\sqrt
s$ energy.

The Coulomb-nuclear interference at the LHC is an ideal laboratory to probe the
exchange of a virtual odd-gluons bound state, because it selects the required
quantum numbers in the $t$-range where the interference terms cannot be
neglected with respect to the QED and nuclear amplitudes. The highest
sensitivity is reached in the $t$-range where the QED and nuclear amplitudes
are of similar magnitude, thus this has been the driving factor in designing
the acceptance requirements then achieved via the $2.5\un{km}$ optics of the
LHC. The $\rho$ parameter being an analytical function of the nuclear phase at
$t=0$, it represents a sensitive probe of the interference terms into the
evolution of the real and imaginary parts of the nuclear amplitude.

Consequently theoretical models have made sensitive predictions via the
evolution of $\rho$ as a function of $\sqrt s$ to quantify the effect of the
possible 3-gluon bound state exchange in the elastic scattering $t$-channel.
Those, currently non-excluded, theoretical models systematically require
significantly lower $\rho$ values at $13\un{TeV}$ than the predicted
Pomeron-only evolution of $\rho$ at $13\un{TeV}$, consistently with the $\rho$
measurement reported in the present work.

The confirmation of this result in additional channels would bring, besides the
evidence for the existence of the QCD-predicted 3-gluon bound state,
theoretical consequences such as the generalization of the Pomeranchuk theorem
(i.e.~the total cross-section of proton-proton and proton-antiproton
asymptotically having their ratio converging to 1 rather than their difference
converging to 0).

On the contrary, if the role of the 3-gluon bound state exchange is shown
insignificant, the present TOTEM results at $13\un{TeV}$ would imply by the
dispersion relations the first experimental evidence for total cross-section
saturation effects at higher energies, eventually deviating from the asymptotic
behaviour proposed by many contemporary models (e.g.~the functional saturation
of the Froissart bound \cite{froissart-1961}).

The two effects, crossing-odd contribution and cross-section saturation, could
both be present without being mutually exclusive.

Besides the extraction of the $\rho$ parameter, the very low $|t|$ elastic
scattering can be used to determine the normalisation of the differential
cross-section -- a crucial ingredient for measurement of the total
cross-section, $\sigma_{\rm tot}$. In its ideal form, the normalisation can be
determined as the proportionality constant between the Coulomb cross-section
known from QED and the data measured at such low $|t|$ that other than Coulomb
cross-section contributions can be neglected. This ``Coulomb normalisation''
technique opens the way to another total cross-section measurement at $\sqrt s
= 13\un{TeV}$, completely independent of previous results. This publication
presents the first successful application of this method to LHC data.

Section~\ref{sec:exp apparatus} of this article outlines the experimental setup
used for the measurement. The properties of the special beam optics are
described in Section~\ref{sec:beam optics}. Section~\ref{sec:data taking} gives
details of the data-taking conditions. The data analysis and reconstruction of
the differential cross-section are described in Section~\ref{sec:differential
cross-section}. Section~\ref{sec:rho} presents the extraction of the $\rho$
parameter and $\sigma_{\rm tot}$ from the differential cross-section. Physics
implications of these new results are discussed in
Section~\ref{sec:discussion}.

\section{Experimental Apparatus}
\label{sec:exp apparatus}

\begin{figure*}
\begin{center}
\includegraphics{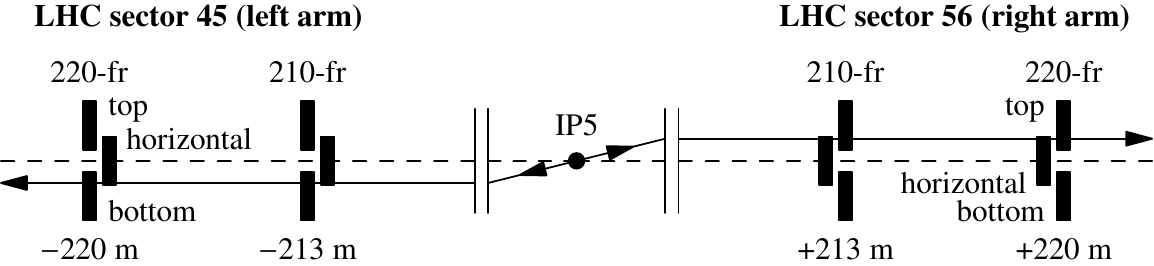}
\caption{%
Schematic view of the RP units used for the presented measurement (black
squares) with two proton tracks from an elastic event (lines with
arrows). The numbers at the bottom indicate the distance from the
interaction point (IP5).
	}
\vskip-6mm
\label{fig:rpsketch}
\end{center}
\end{figure*}

The TOTEM experiment, located at the LHC Interaction Point (IP) 5 together with
the CMS experiment, is dedicated to the measurement of the total cross-section,
elastic scattering and diffractive processes. The experimental apparatus,
symmetric with respect to the IP, detects particles at different scattering
angles in the forward region: a forward proton spectrometer composed of
detectors in Roman Pots (RPs) and the magnetic elements of the LHC and, to
measure at larger angles, the forward tracking telescopes T1 and T2. A complete
description of the TOTEM detector instrumentation and its performance is given
in~\cite{totem-jinst} and~\cite{totem-ijmp}. The data analysed here come from
the RPs only. 

A RP is a movable beam-pipe insertion which houses the tracking detectors that
are thus capable of approaching the LHC beam to a distance of less than a
millimetre, and to detect protons with scattering angles of only a few
microradians. The proton spectrometer is organised in two arms: one on the left
side of the IP (LHC sector 45) and one on the right (LHC sector 56), see
Figure~\ref{fig:rpsketch}. In each arm, there are several RP units. The
presented measurement is performed with units ``210-fr'' (approximately
$213\un{m}$ from the IP) and ``220-fr'' (about $220\un{m}$ from the IP). The
210-fr unit is tilted by $8^\circ$ in the transverse plane with respect to the
220-fr unit. Each unit consists of 3 RPs, one approaching the outgoing beam
from the top, one from the bottom, and one horizontally. Each RP houses a stack
of 5 ``U'' and 5 ``V'' silicon strip detectors, where ``U'' and ``V'' refer to
two mutually perpendicular strip orientations. The special design of the
sensors is such that the insensitive area at the edge facing the beam is only a
few tens of micrometres \cite{edgeless-strips}. Due to the $7\un{m}$ long lever
arm between the two RP units in one arm, the local track angles can be
reconstructed with an accuracy  of about $2.5\un{\mu rad}$.

Since elastic scattering events consist of two collinear protons emitted in
opposite directions, the detected events can have two topologies, called
``diagonals'': 45 bottom -- 56 top and 45 top -- 56 bottom, where the numbers
refer to the LHC sector.

This article uses a reference frame where $x$ denotes the horizontal axis
(pointing out of the LHC ring), $y$ the vertical axis (pointing against
gravity) and $z$ the beam axis (in the clockwise direction).

\section{Beam Optics}
\label{sec:beam optics}

The beam optics relate the proton kinematical states at the IP and at the RP
location. A proton emerging from the interaction vertex $(x^*$, $y^*)$ at the
angle $(\theta_x^*,\theta_y^*)$ (relative to the $z$ axis) and with momentum
$p\,(1+\xi)$, where $p$ is the nominal initial-state proton momentum, is
transported along the outgoing beam through the LHC magnets. It arrives at the
RPs in the transverse position
\begin{equation}
\label{eq:prot trans}
	\begin{aligned}
		x(z_{\rm RP}) =& L_x(z_{\rm RP})\, \theta_x^*\ +\ v_x(z_{\rm RP})\, x^*\ +\ D_x(z_{\rm RP})\, \xi\ ,\cr
		y(z_{\rm RP}) =& L_y(z_{\rm RP})\, \theta_y^*\ +\ v_y(z_{\rm RP})\, y^*\ +\ D_y(z_{\rm RP})\, \xi \quad
	\end{aligned}
\end{equation}

relative to the beam centre. This position is determined by the optical
functions, characterising the transport of protons in the beam line and
controlled via the LHC magnet currents.  The effective length $L_{x,y}(z)$, the
magnification $v_{x,y}(z)$ and the dispersion $D_{x,y}(z)$ quantify the
sensitivity of the measured proton position to the scattering angle, the vertex
position and the momentum loss, respectively. Note that for elastic collisions
the dispersion terms $D\,\xi$ can be ignored because the protons do not lose
any momentum. The values of $\xi$ only account for the initial state momentum
offset ($\approx 10^{-3}$) and variations ($\approx 10^{-4}$). Due to the
collinearity of the two elastically scattered protons and the symmetry of the
optics, the impact of $D\,\xi$ on the reconstructed scattering angles is
negligible compared to other uncertainties.


The data for the analysis presented here have been taken with a new, special
optics, the $\beta^{*} = 2500\un{m}$, specifically developed for measuring
low-$|t|$ elastic scattering and conventionally labelled by the value of the
$\beta$-function at the interaction point. It maximises the vertical effective
length $L_{y}$ and minimises the vertical magnification $|v_{y}|$ at the RP
position $z = 220\,$m (Table~\ref{tab:optics}). This configuration is called
``parallel-to-point focussing'' because all protons with the same angle in the
IP are focussed on one point in the RP at 220\,m. It optimises the sensitivity
to the vertical projection of the scattering angle -- and hence to $|t|$ --
while minimising the influence of the vertex position. In the horizontal
projection the parallel-to-point focussing condition is not fulfilled, but --
similarly to the $\beta^{*} = 1000\,$m optics used for a previous
measurement~\cite{totem-8tev-1km} -- the effective length $L_{x}$ at $z =
220\,$m is sizeable, which reduces the uncertainty in the horizontal component
of the scattering angle. The very high value of $\beta^*$ also implies very low
beam divergence which is essential for accurate measurement at very low $|t|$.

\begin{table}
\caption{
	Optical functions for elastic proton transport for the $\beta^{*} =
	2500\,$m optics. The values refer to the right arm, for the left one
	they are very similar.
}
\label{tab:optics}
\begin{center}
\vskip-3mm
\begin{tabular}{ccccc}\hline
RP unit & $L_x$ & $v_x$ & $L_y$ & $v_y$ \cr\hline
210-fr & $73.05\un{m}$ & $-0.634$ & $244.68\un{m}$ & $+0.009$ \cr
220-fr & $51.10\un{m}$ & $-0.540$ & $282.96\un{m}$ & $-0.018$ \cr
\hline
\end{tabular}
\end{center}
\end{table}

\section{Data Taking}
\label{sec:data taking}

The results reported here are based on data taken in September 2016 during a
sequence of dedicated LHC proton fills (5313, 5314, 5317 and 5321) with the
special beam properties described in the previous section.



The vertical RPs approached the beam centre to only about 3 times the vertical
beam width, $\sigma_{y}$, thus roughly to $0.4\un{mm}$. The exceptionally close
distance was required in order to reach very low $|t|$ values and was possible
due to the low beam intensity in this special beam operation: each beam
contained only four or five colliding bunches and one non-colliding bunch, each
with about $5\times 10^{10}$ protons.

The horizontal RPs were only needed for the track-based alignment and therefore
placed at a safe distance of $8\,\sigma_{x} \approx 5$\,mm, close enough to
have an overlap with the vertical RPs.

\begin{figure*}
\begin{center}
\includegraphics{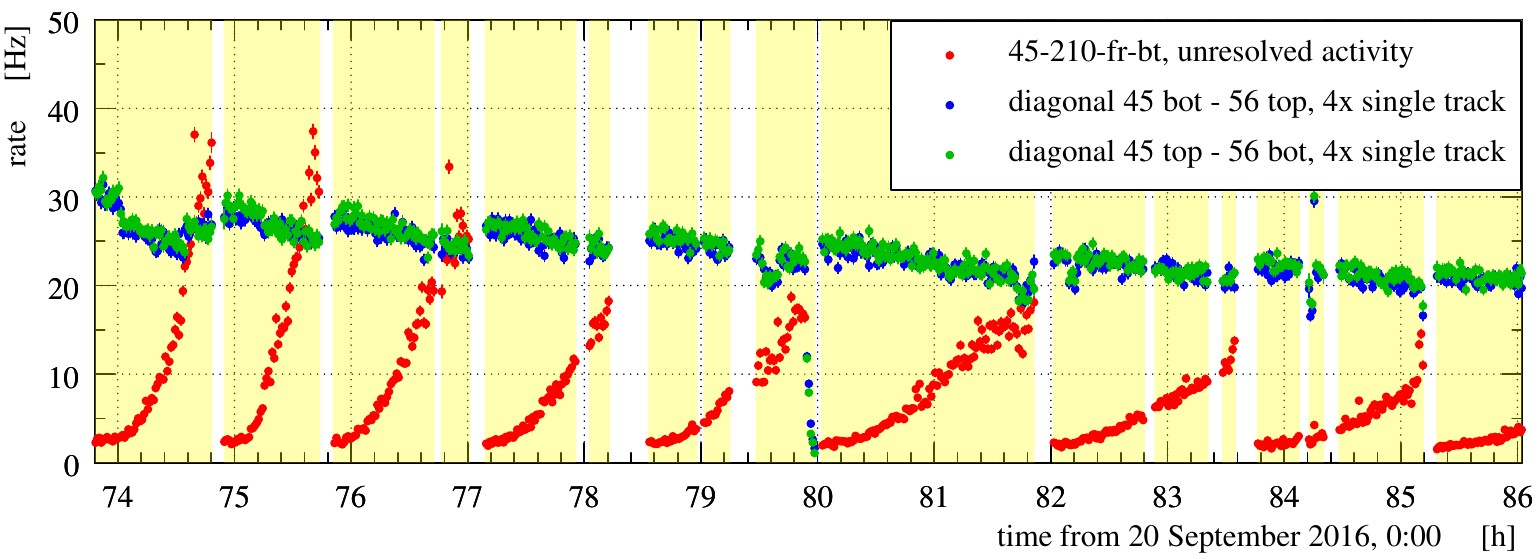}
\caption{%
	Event rates from run 5321 as a function of time. The blue and green
	graphs give rates of fully reconstructed events in the two diagonal
	configurations relevant for elastic scattering. The red graph shows a
	rate of high-multiplicity events in a single RP (bottom pot in sector
	45 and unit 210-fr) where no track can be reconstructed. For the other
	RPs this rate evolution was similar.
	}
\label{fig:rates_vs_time}
\end{center}
\end{figure*}

The collimation strategy applied in the previous measurement
\cite{totem-8tev-1km} with carbon primary collimators was first tried, however,
this resulted in too high beam halo background. To keep the background under
control, a new collimation scheme was developed, with more absorbing tungsten
collimators closest to the beam in the vertical plane, in order to minimise the
out-scattering of halo particles. As a first step, vertical collimators TCLA
scraped the beam down to $2\,\sigma_{y}$, then the collimators were retracted
to $2.5\,\sigma_{y}$, thus creating a $0.5\,\sigma_{y}$ gap between the beam
edge and the collimator jaws. A similar procedure was performed in the
horizontal plane: collimators TCP.C scraped the beam to $3\un{\sigma_{x}}$ and
then were retracted to $5.5\un{\sigma_{x}}$, creating a $2.5\un{\sigma_x}$ gap.
With the halo strongly suppressed and no collimator producing showers by
touching the beam, the RPs at $3\,\sigma_{y}$ were operated in a
background-depleted environment for about one hour until the beam-to-collimator
gap was refilled by diffusion, as diagnosed by the increasing shower rate (red
graph in Figure~\ref{fig:rates_vs_time}). When the background conditions had
deteriorated to an unacceptable level, the beam cleaning procedure was
repeated, again followed by a quiet data-taking period.

The events collected were triggered by a double-arm proton trigger (coincidence
of any RP left of IP5 and any RP right of IP5) or a zero-bias trigger (random
bunch crossings) for calibration purposes.

In total, a data sample with an integrated luminosity of about $0.4\,\rm
nb^{-1}$ was accumulated in which more than 7 million of elastic event
candidates were tagged.

\section{Differential Cross-Section}
\label{sec:differential cross-section}

The analysis method is very similar to the previously published one
\cite{totem-8tev-1km}. The only important difference stems from using different
RPs for the measurement: unit 210-fr instead of 220-nr as in
\cite{totem-8tev-1km} since the latter was not equipped with sensors anymore.
Due to the optics and beam parameters the unit 210-fr has worse low-$|t|$
acceptance, further deteriorated by the tilt of the unit (effectively
increasing the RP distance from the beam). Consequently, in order to maintain
the low-$|t|$ reach essential for this study, the main analysis (denoted
``2RP'') only uses the 220-fr units (thus 2 RPs per diagonal). Since not using
the 210-nr units may, in principle, result in worse resolution and background
suppression, for control reasons, the traditional analysis with 4 units per
diagonal (denoted ``4RP'') was pursued, too. In Section~\ref{sec:cross checks}
the ``2RP'' and ``4RP'' will be compared showing a very good agreement. In what
follows, the ``2RP'' analysis will be described unless stated otherwise.

Section~\ref{sec:event analysis} covers all aspects related to the
reconstruction of a single event. Section~\ref{sec:diff cs} describes the steps
of transforming a raw $t$-distribution into the differential cross-section. The
$t$-distributions are analysed separately for each LHC fill and each diagonal,
and are only merged at the end as detailed in Section~\ref{sec:data merging}.
Section~\ref{sec:systematics} describes the evaluation of systematic
uncertainties and Section~\ref{sec:cross checks} presents several comparison
plots used as systematic cross checks.

\subsection{Event Analysis}
\label{sec:event analysis}

The event kinematics are determined from the coordinates of track hits in the
RPs after proper alignment (see Sec.~\ref{sec:alignment}) using the LHC optics
(see Sec.~\ref{sec:optics}).


\subsubsection{Kinematics Reconstruction}
\label{sec:kinematics}

For each event candidate the scattering angles of both protons (one per arm)
are first estimated separately. In the ``2RP'' analysis, these formulae are
used:
\begin{equation}
\label{eq:kin 1a}
	\theta^{*\rm L,R}_x = {x\over L_x}\ ,\quad \theta^{*\rm L,R}_y = {y\over L_y}
\end{equation}
where L and R refer to the left and right arm, respectively, and $x$ and $y$
stand for the proton position in the 220-fr unit. This one-arm reconstruction
is used for tagging of elastic events, where the left and right arm protons are
compared.

Once a proton pair has been selected, both arms are used to reconstruct the
kinematics of the event
\begin{equation}
\label{eq:kin 2a}
		\theta_x^* = {1\over 2} \left( \theta^{*\rm L}_x + \theta^{*\rm R}_x \right)\ ,\qquad
		\theta_y^* = {1\over 2} \left( \theta^{*\rm L}_y + \theta^{*\rm R}_y \right)\ .
\end{equation}
Thanks to the left-right symmetry of the optics and elastic events, this
combination leads to cancellation of the vertex terms (cf.~Eq.~(\ref{eq:prot
trans})) and thus to improvement of the angular resolution (see Section
\ref{sec:resolution}).

Finally, the scattering angle, $\theta^*$, and the four-momentum transfer
squared, $t$, are calculated:

\begin{equation}
\label{eq:th t}
\theta^* = \sqrt{{\theta_x^*}^2 + {\theta_y^*}^2}\ ,\qquad t = - p^2 ({\theta_x^*}^2 + {\theta_y^*}^2)\ ,
\end{equation}
where $p$ denotes the beam momentum.

In the ``4RP'' analysis, the same reconstruction as in \cite{totem-8tev-1km} is
used which allows for stronger elastic-selection cuts, see
Section~\ref{sec:tagging}.


\subsubsection{Alignment}
\label{sec:alignment}

TOTEM's usual three-stage procedure (Section 3.4 in~\cite{totem-ijmp}) for
correcting the detector positions and rotation angles has been applied: a
beam-based alignment prior to the run followed by two offline methods. The
first method uses straight tracks to determine the relative position among the
RPs by minimising track-hit residuals. The second method exploits the
symmetries of elastic scattering to determine the positions of RPs with respect
to the beam. This determination is repeated in 20-minute time intervals to
check for possible beam movements.

The alignment uncertainties have been estimated as $25\un{\mu m}$ (horizontal
shift), $100\un{\mu m}$ (vertical shift) and $2\un{m rad}$ (rotation about the
beam axis). Propagating them through Eq.~(\ref{eq:kin 2a}) to reconstructed
scattering angles yields $0.50\un{\mu rad}$ ($0.35\un{\mu rad}$) for the
horizontal (vertical) angle. RP rotations induce a bias in the reconstructed
scattering angles:
\begin{equation}
\label{eq:alig rot bias}
	\theta_x^* \rightarrow \theta_x^* + c \theta_y^*\ ,\quad
	\theta_y^* \rightarrow \theta_y^* + d \theta_x^*\ ,
\end{equation}
where the proportionality constants $c$ and $d$ have zero mean and standard
deviations of $0.013$ and $0.00039$, respectively.


\subsubsection{Optics}
\label{sec:optics}

It is crucial to know with high precision the LHC beam optics between IP5 and
the RPs, i.e. the behaviour of the spectrometer composed of the various
magnetic elements. The optics calibration has been applied as described
in~\cite{totem-optics}. This method uses RP observables to determine fine
corrections to the optical functions presented in Eq.~(\ref{eq:prot trans}).

In each arm, the residual errors induce a bias in the reconstructed scattering angles:
\begin{equation}
\label{eq:opt bias}
	\theta_x^* \rightarrow (1 + b_x)\, \theta_x^*\ ,\qquad
	\theta_y^* \rightarrow (1 + b_y)\, \theta_y^*\ ,
\end{equation}
where the biases $b_x$ and $b_y$ have uncertainties of $0.17\un{\%}$ and
$0.15\un{\%}$, respectively, and a correlation factor of $-0.90$. To evaluate
the impact on the $t$-distribution, it is convenient to decompose the
correlated biases $b_x$ and $b_y$ into eigenvectors of the covariance matrix:
\begin{equation}
\label{eq:opt bias modes}
\begin{pmatrix} b_x^{\rm L}\cr b_y^{\rm L} \cr b_x^{\rm R}\cr b_y^{\rm R} \end{pmatrix} =
	   \eta_1 \underbrace{\begin{pmatrix} -1.608\times10^{-3}\cr +1.473\times10^{-3}\cr -1.630\times10^{-3}\cr +1.477\times10^{-3} \end{pmatrix}}_{\rm mode\ 1}
  \ +\ \eta_2 \underbrace{\begin{pmatrix} -5.157\times10^{-4}\cr +2.541\times10^{-5}\cr +5.566\times10^{-4}\cr +2.746\times10^{-5} \end{pmatrix}}_{\rm mode\ 2}
  \ +\ \eta_3 \underbrace{\begin{pmatrix} +3.617\times10^{-4}\cr +3.625\times10^{-4}\cr +3.006\times10^{-4}\cr +3.641\times10^{-4} \end{pmatrix}}_{\rm mode\ 3}\ ,
\end{equation}
where the factors $\eta_{1,2,3}$ have zero mean and unit variance. The fourth
eigenmode has a negligible contribution and therefore is not explicitly listed.


\subsubsection{Resolution}
\label{sec:resolution}

Two kinds of resolution can be distinguished: the resolution of the single-arm
angular reconstruction, Eq.~(\ref{eq:kin 1a}), used for selection cuts and
near-edge acceptance correction, and the resolution of the double-arm
reconstruction, Eq.~(\ref{eq:kin 2a}), used for the unsmearing correction of
the final $t$-distribution. Since the single-arm reconstruction is biased by
the vertex term in the horizontal plane, the corresponding resolution is
significantly worse than the double-arm reconstruction.

\begin{figure}
\begin{center}
\includegraphics{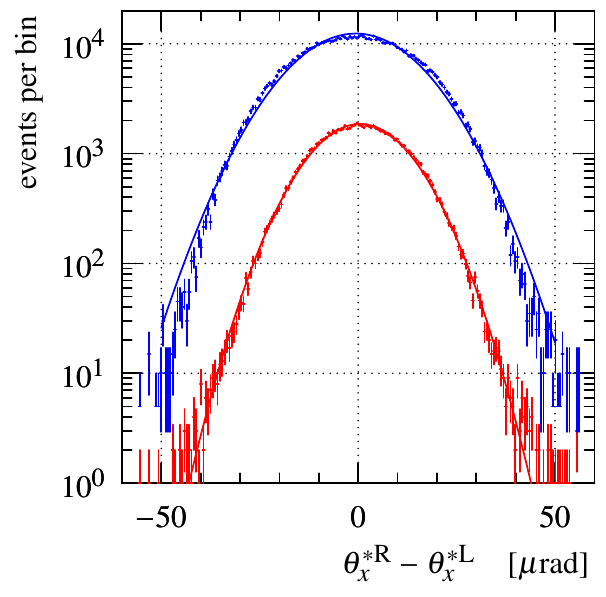}
\caption{%
	Difference between horizontal scattering angles reconstructed in the
	right and left arm, for the diagonal 45 bottom - 56 top. Red: data from
	the beginning of fill 5317, blue: data from the fill end (vertically
	scaled $5\times$). The solid lines represent Gaussian fits.
	}
\label{fig:resol x 1a}
\end{center}
\end{figure}

The single-arm resolution can be studied by comparing the angles reconstructed
from the left and right arm, see an example in Figure~\ref{fig:resol x 1a}. The
width of the distributions was found to grow slightly during the fills,
compatible with the effect of beam emittance growth. The typical range was from
$10.0$ to $14.5\un{\mu rad}$ for the horizontal projection and from $0.36$ to
$0.38\un{\mu rad}$ for the vertical. The associated uncertainties were $0.3$
and $0.007$, respectively. As illustrated in Figure~\ref{fig:resol x 1a}, the
shape of the distributions is very close to Gaussian, especially at the
beginning of each fill.

Since in the vertical plane the resolution is driven by the beam divergence,
the double-arm resolution can simply be scaled from the single-arm value:
$\sigma(\theta^*_y) = (0.185 \pm 0.010)\un{\mu rad}$ where the uncertainty
accounts for the full variation in time. In the horizontal plane the estimation
is more complex due to several contributing smearing mechanisms. Therefore, a
MC study was performed with two extreme sets of beam divergence, vertex size
and sensor resolution values. These parameters were tuned within the ``4RP''
analysis where they are accessible thanks to the additional information from
the 210-fr units. The study yielded $\sigma(\theta^*_x) =  (0.29 \pm
0.04)\un{\mu rad}$ where the uncertainty accounts for the full time variation.

\subsection{Differential Cross-Section Reconstruction}
\label{sec:diff cs}

For a given $t$ bin, the differential cross-section is evaluated by selecting
and counting elastic events:
\begin{equation}
{\d\sigma\over \d t}(\hbox{bin}) =
	\mathcal{N}\, \mathcal{U}(t)\, \mathcal{B}(t)\, {1\over \Delta t}
	\sum\limits_{\vbox{\scriptsize\hbox{\hskip0.7mm events}\vskip-1mm\hbox{$t\, \in\, {\rm bin}$}}} \mathcal{A}(\theta^*_x, \theta_y^*)\ \mathcal{E}(\theta_y^*)
	\ ,
\end{equation}
where $\Delta t$ is the width of the bin, $\mathcal{N}$ is a normalisation
factor and the other symbols stand for various correction factors:
$\mathcal{U}$ for unfolding of resolution effects, $\mathcal{B}$ for background
subtraction, $\mathcal{A}$ for acceptance correction and $\mathcal{E}$ for
detection and reconstruction efficiency.


\subsubsection{Event Tagging}
\label{sec:tagging}

\begin{table}
\caption{
	The elastic selection cuts. The superscripts R and L refer to the right
	and left arm. The rightmost column gives a typical standard deviation
	of the cut distribution.
}
\label{tab:cuts}
\begin{center}
\begin{tabular}{ccc}\hline
number & cut & std.~dev.~($\equiv 1\sigma$)\cr\hline
\vrule width0pt height11pt
1 & $\theta_x^{*\rm R} - \theta_x^{*\rm L}$				& $14\un{\mu rad}$	\cr
2 & $\theta_y^{*\rm R} - \theta_y^{*\rm L}$				& $0.38\un{\mu rad}$	\cr\hline
\end{tabular}
\end{center}
\end{table}

\begin{figure}
\begin{center}
\includegraphics{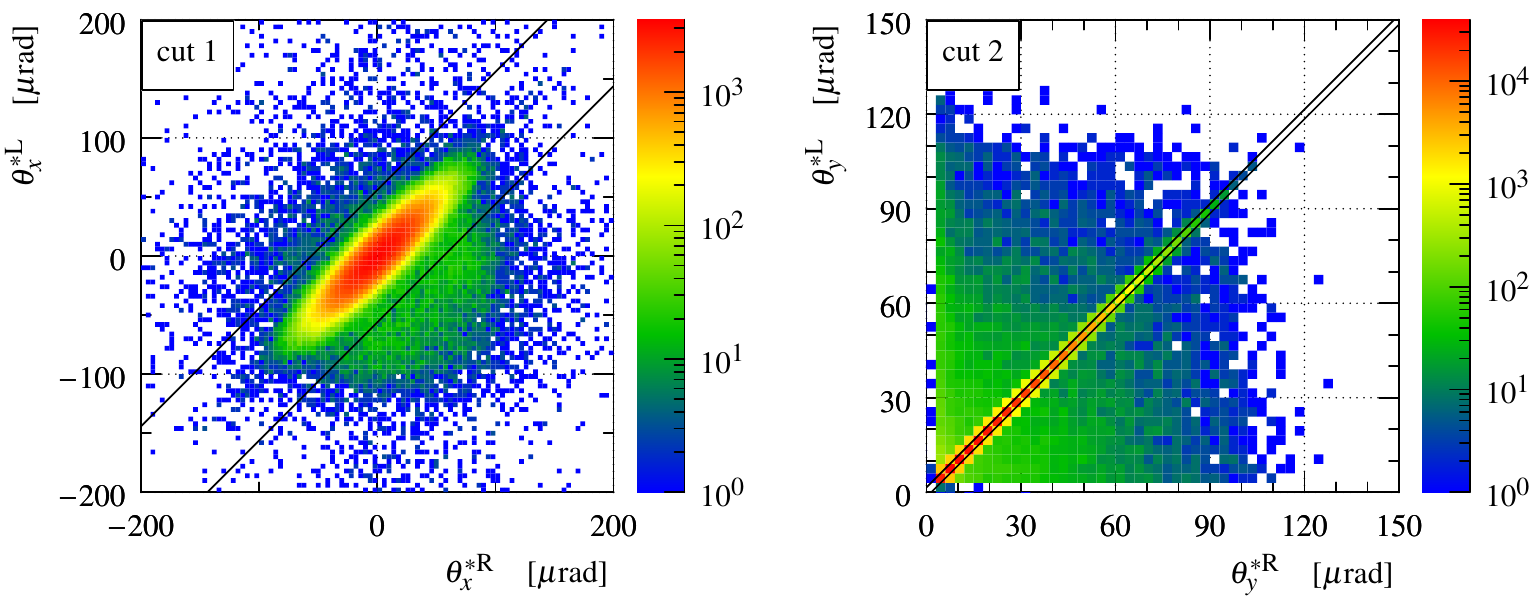}
\caption{%
	Correlation plots for the elastic event selection cuts presented in
	Table~\ref{tab:cuts} (``2RP'' analysis), showing events from the LHC
	fill 5313 and with diagonal topology 45 bottom -- 56 top. The solid
	black lines delimit the signal region within $\pm 4\un{\sigma}$.
	}
\label{fig:cuts}
\end{center}
\end{figure}

Within the ``2RP'' analysis one may apply the cuts requiring the
reconstructed-track collinearity between the left and the right arm, see
Table~\ref{tab:cuts}. The correlation plots corresponding to these cuts are
shown in Figure~\ref{fig:cuts}. 

In order to limit the selection inefficiency, the thresholds for the cuts are
set to $4\un{\sigma}$. Applying the cuts at the $5\un{\sigma}$-level would
yield about $0.1\un{\%}$ more events almost uniformly in every $|t|$-bin. This
kind of inefficiency only contributes to a global scale factor, which is
irrelevant for this analysis because the normalisation is taken from a
different data set (cf. Section~\ref{sec:normalisation}).

In the ``4RP'' analysis, thanks to the additional information from the 210-fr units, more cuts can be applied (cf.~Table~2 in~\cite{totem-7tev-el}). In particular the left-right comparison of the reconstructed horizontal vertex position, $x^*$, and the vertical position-angle correlation in each arm. Furthermore, since the single-arm reconstruction can disentangle the contributions from $x^*$ and $\theta^*_x$, the angular resolution is better compared with the ``2RP'' analysis and consequently cut 1 in the ``4RP'' analysis is more efficient against background.


\subsubsection{Background}
\label{sec:background}

As the RPs were very close to the beam, one may expect an enhanced background
from coincidence of beam halo protons hitting detectors in the two arms. Other
background sources (pertinent to any elastic analysis) are central diffraction
and pile-up of two single diffraction events.

The background rate (i.e.~impurity of the elastic tagging) is estimated in two
steps, both based on distributions of discriminators from Table~\ref{tab:cuts}
plotted in various situations, see an example in Figure~\ref{fig:tag bckg
integ}. In the first step, diagonal data are studied under several cut
combinations. While the central part (signal) remains essentially constant, the
tails (background) are suppressed when the number of cuts is increased. In the
second step, the background distribution is interpolated from the tails into
the signal region. The form of the interpolation is inferred from non-diagonal
RP track configurations (\textit{45 bottom -- 56 bottom} or \textit{45 top --
56 top}), artificially treated like diagonal signatures by inverting the $y$
coordinate sign in the arm 45. These non-diagonal configurations cannot contain
any elastic signal and hence consist purely of background which is expected to
be similar in the diagonal and non-diagonal configurations. This expectation is
supported by the agreement of the tails of the red, blue and green curves in
the figure. Since the non-diagonal distributions are flat, the comparison of
the signal-peak size to the amount of interpolated background yields an
order-of-magnitue estimate of $1 - \mathcal{B} = \mathcal{O}(10^{-3})$.

\begin{figure}
\begin{center}
\includegraphics{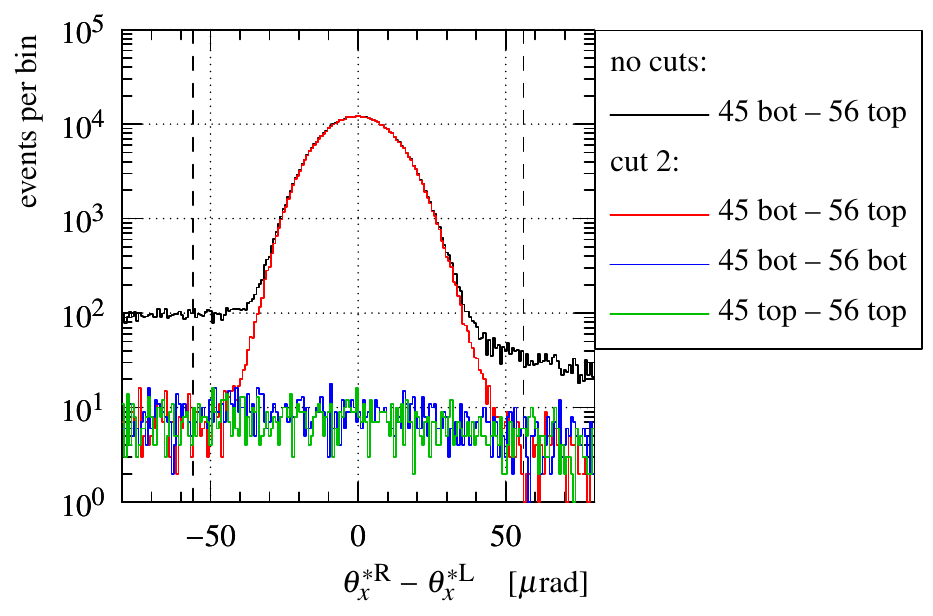}
\caption{%
	Distributions of discriminator 1, i.e. the difference between the
	horizontal scattering angle reconstructed from the right and the left
	arm. Data from LHC fill 5314. Black and red curves: data from diagonal
	45 bottom -- 56 top, the different colours correspond to various
	combinations of the selection cuts (see numbering in
	Table~\ref{tab:cuts}). Blue and green curves: data from anti-diagonal
	RP configurations, obtained by inverting track $y$ coordinate in the
	left arm. The vertical dashed lines represent the boundaries of the
	signal region ($\pm 4\un{\sigma}$).
	}
\label{fig:tag bckg integ}
\end{center}
\end{figure}

\begin{figure}
\begin{center}
\includegraphics{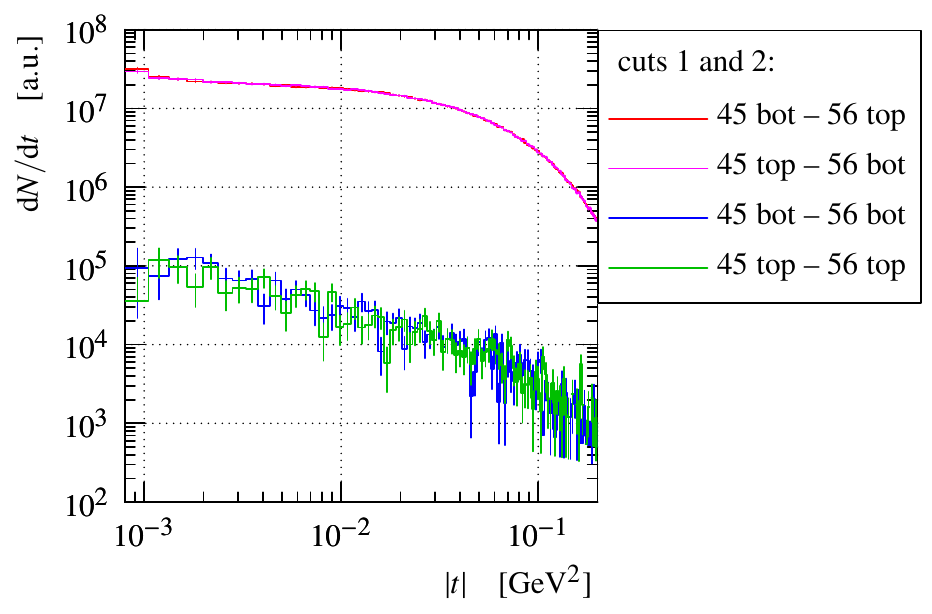}
\caption{%
	Comparison of $|t|$-distributions from different diagonal (signal +
	background) and anti-diagonal (background) configurations, after all
	cuts and acceptance correction. Data from the LHC fill 5314.
	}
\label{fig:tag bckg dist}
\end{center}
\end{figure}

The $t$-distribution of the background can also be estimated by comparing data
from diagonal and anti-diagonal configurations, as illustrated in
Figure~\ref{fig:tag bckg dist}. The ratio background / (signal + background)
can be obtained by dividing the blue or green histograms by the red or magenta
histograms. Consequently, the background correction factor, $\mathcal{B}$, is
estimated to be $0.9975 \pm 0.0010$ at $|t| = 0.001\un{GeV^2}$, $0.9992 \pm
0.0003$ at $|t| = 0.05\un{GeV^2}$ and $0.998 \pm 0.001$ at $|t| =
0.2\un{GeV^2}$. The uncertainty comes from statistical fluctuations in the
histograms and from considering different diagonals and anti-diagonals.

\subsubsection{Acceptance Correction}
\label{sec:acc corr}

\begin{figure}
\begin{center}
\includegraphics{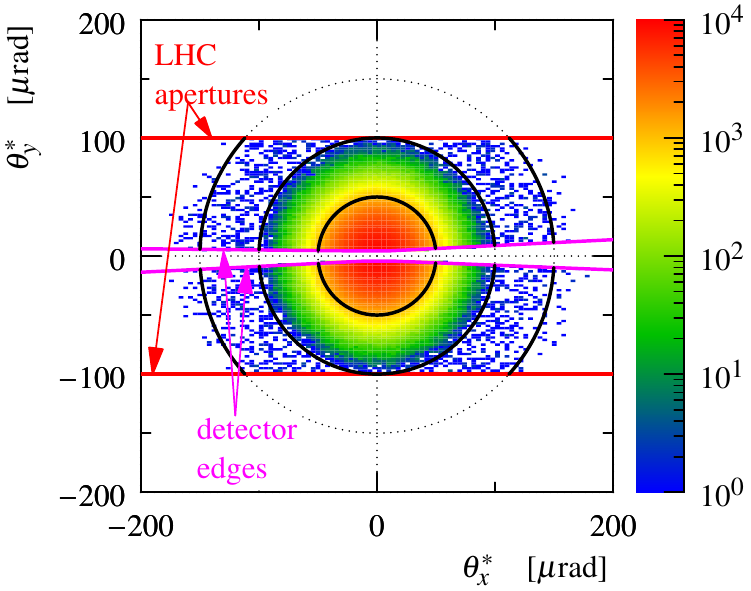}
\caption{%
	Distribution of scattering angle projections $\theta_y^*$
	vs.~$\theta_x^*$, data from LHC fill 5317. The upper (lower) part comes
	from the diagonal 45 bottom -- 56 top (45 top -- 56 bottom). The red
	horizontal lines represent cuts due to the LHC apertures, the magenta
	lines cuts due to the RP edges. The dotted circles show contours of
	constant scattering angle $\theta^* = 50$, $100$ and $150\un{\mu rad}$.
	The parts of the contours within acceptance are emphasized in thick
	black.
	}
\label{fig:acc corr princ}
\end{center}
\end{figure}

The acceptance for elastic protons is limited mostly by two factors: sensor
coverage (relevant for low $|\theta^*_y|$) and LHC beam aperture (at
$|\theta^*_y| \approx 100\un{\mu rad}$). Since the 210-fr unit is tilted with
respect to the 220-fr unit, the thin windows around sensors do not overlap
perfectly. Therefore there are phase space regions where protons need to
traverse thick walls of 210-fr RP before being detected in 220-fr RP. This
induces reduced detection efficiency difficult to determine precisely.
Consequently these regions (close to the sensor edge facing the beam) have been
excluded from the fiducial region used in the analysis, see the magenta lines
in Figure~\ref{fig:acc corr princ}.

The correction for the above phase-space limitations includes two contributions
-- a geometrical correction $\mathcal{A}_{\rm geom}$ reflecting the fraction of
the phase space within the acceptance and a component $\mathcal{A}_{\rm fluct}$
correcting for fluctuations around the acceptance boundaries:
\begin{equation}
\mathcal{A}(\theta^*_x, \theta_y^*) = \mathcal{A}_{\rm geom}(\theta^*)\ \mathcal{A}_{\rm fluct}(\theta^*_x, \theta_y^*)\ .
\end{equation}

The calculation of the geometrical correction $\mathcal{A}_{\rm geom}$ is based
on the azimuthal symmetry of elastic scattering, experimentally verified for
the data within acceptance. As shown in Figure \ref{fig:acc corr princ}, for a
given value of $\theta^*$ the correction is given by:
\begin{equation}
\label{eq:acc geom}
\mathcal{A_{\rm geom}}(\theta^*) = {
	\hbox{full circumference}\over 
	\hbox{arc length within acceptance}
} \ .
\end{equation}

The correction $\mathcal{A}_{\rm fluct}$ is calculated analytically from the
probability that any of the two elastic protons leaves the region of acceptance
due to the beam divergence. The beam divergence distribution is modelled as a
Gaussian with the spread determined by the method described in
Section~\ref{sec:resolution}. This contribution is sizeable only close to the
acceptance limitations. Data from regions with corrections larger than $2$ are
discarded.

The full acceptance correction, $\mathcal{A}$, has a value of $12$ in the
lowest-$|t|$ bin and decreases smoothly towards about $2.1$ at $|t| =
0.2\un{GeV^2}$. Since a single diagonal cannot cover more than half of the
phase space, the minimum value of the correction is $2$.

The uncertainties related to $\mathcal{A}_{\rm fluct}$ follow from the
uncertainties of the resolution parameters: standard deviation and distribution
shape, see Section~\ref{sec:resolution}. Since $\mathcal{A}_{\rm geom}$ is
calculated from a trivial trigonometric formula, there is no uncertainty
directly associated with it. However biases can arise indirectly from effects
that break the assumed azimuthal symmetry like misalignments or optics
perturbations already covered above.


\subsubsection{Inefficiency Corrections}
\label{sec:ineff corr}

Since the overall normalisation will be determined from another dataset (see
Section~\ref{sec:normalisation}), any inefficiency correction that does not
alter the $t$-distribution shape does not need to be considered in this
analysis (trigger, data acquisition and pile-up inefficiency discussed
in~\cite{totem-7tev-el,totem-8tev-tot1}). The remaining inefficiencies are
related to the inability of a RP to resolve the elastic proton track.

\begin{figure}
\begin{center}
\includegraphics{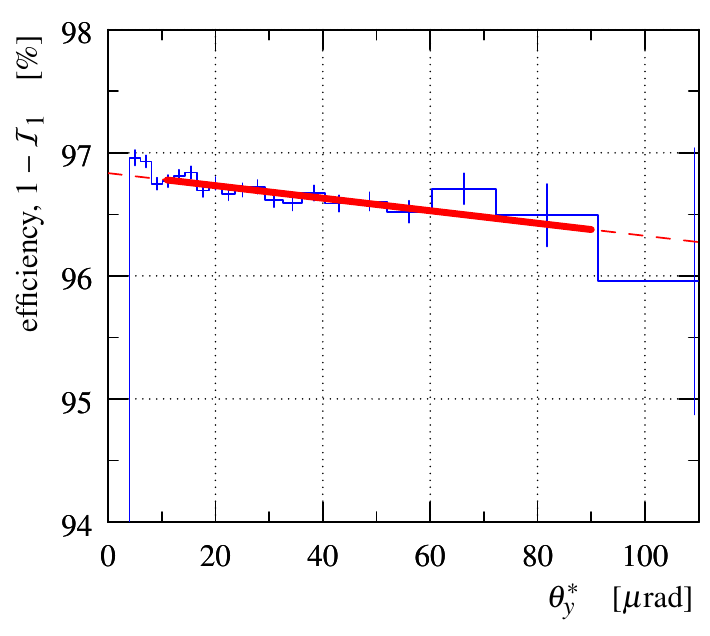}
\caption{%
	Single-RP uncorrelated inefficiency for the 220-fr bottom RP in the
	right arm. The rapid drop at $\theta_y^* \approx 4\un{\mu rad}$ is due
	to acceptance effects at the sensor edge. The red lines represent a
	linear fit of the efficiency dependence on the vertical scattering
	angle (solid) and its extrapolation to the regions affected by
	acceptance effects (dashed).
	}
\label{fig:eff uncorr}
\end{center}
\end{figure}

One such case is when a single RP does not detect and/or reconstruct a proton
track, with no correlation to other RPs. This type of inefficiency,
$\mathcal{I}_1$, is evaluated within the ``4RP'' analysis by removing the
studied RP from the tagging cuts, repeating the event selection and calculating
the fraction of recovered events. A typical example is given in
Figure~\ref{fig:eff uncorr}, showing that the efficiency decreases gently with
the vertical scattering angle. This dependence originates from the fact that
protons with larger $|\theta_y^*|$ hit the RPs further from their edge and
therefore the potentially created secondary particles have more chance to be
detected. Since the RP detectors cannot resolve multiple tracks (non-unique
association between ``U'' and ``V'' track candidates), the presence of a
secondary particle track prevents from using the affected RP in the analysis.

Proton interactions in a RP affecting simultaneously another RP downstream
represent another source of inefficiency. The contribution from these
correlated inefficiencies, $\mathcal{I}_2$, is determined by evaluating the
rate of events with high track multiplicity ($\gtrsim$ 5) in both 210-fr and
220-fr RP units. Events with high track multiplicity simultaneously in the top
and bottom RP of the 210-fr units are discarded as such a shower is likely to
have started upstream from the RP station and thus be unrelated to the elastic
proton interacting with detectors. The value, $\mathcal{I}_2 \approx (1.5 \pm
0.7)\un{\%}$, is compatible between left/right arms and top/bottom RP pairs and
compares well to Monte-Carlo simulations (e.g.~section 7.5 in
\cite{hubert-thesis}).

The full correction is calculated as
\begin{equation}
\label{efficiency}
	\mathcal{E}(\theta_y^*) = {1\over 1 - \left( \sum\limits_{i\in \rm RPs} \mathcal{I}^i_1(\theta_y^*) + 2 \mathcal{I}_2 \right) } \ .
\end{equation}

The first term in the parentheses sums the contributions from the diagonal RPs
used in the analysis. In the ``2RP'' analysis it increases from about $6.9$ to
$8.5\un{\%}$ from the lowest to the highest $|\theta_y^*|$, with an uncertainty
of about $0.4\un{\%}$. For the ``4RP'' analysis, since more RPs contribute, the
sum is greater: from $10.5$ to $13.0\un{\%}$ between the lowest to the highest
$|\theta_y^*|$.


\subsubsection{Unfolding of Resolution Effects}
\label{sec:unfolding}

\begin{figure}
\begin{center}
\includegraphics{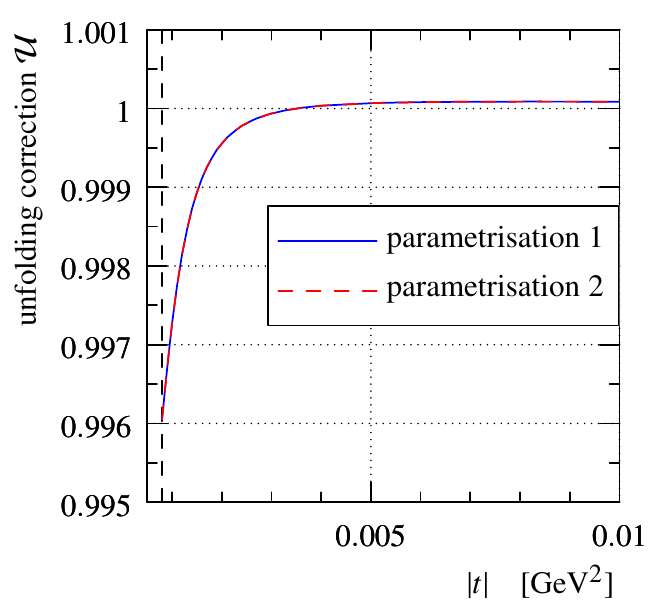}
\caption{%
	Unfolding correction as a function of $|t|$. The vertical dashed line
	indicates the position of the acceptance cut. The two correction curves
	were obtained with different fit parametrisations used in step 1 (see
	text).
	}
\label{fig:unfolding}
\end{center}
\end{figure}

Thanks to the very good resolution (see Section~\ref{sec:resolution}), the
following iterative procedure can be safely used to evaluate the correction for
resolution effects.
\begin{itemize}
\item[1.] The differential cross-section data are fitted by a smooth curve.
\item[2.] The fit is used in a numerical-integration calculation of the smeared $t$-distribution (using the resolution parameters determined in Section~\ref{sec:resolution}). The ratio between the smeared and the non-smeared $t$-distributions gives a set of per-bin correction factors.
\item[3.] The corrections are applied to the observed (yet uncorrected) differential cross-section yielding a better estimate of the true $t$-distribution.
\item[4.] The corrected differential cross-section is fed back to step 1.
\end{itemize}
As the estimate of the true $t$-distribution improves, the difference between
the correction factors obtained in two successive iterations decreases. When
the difference becomes negligible, the iteration stops. This is typically
achieved after the second iteration. 

The final correction $\mathcal{U}$ is significantly different from $1$ only at
very low $|t|$ (where a rapid cross-section growth occurs, see
Figure~\ref{fig:unfolding}). The relative effect is never greater than
$0.4\un{\%}$.

Several fit parametrisations were tested, however yielding negligible
difference in the final correction $\mathcal{U}$ for $|t| \lesssim
0.3\un{GeV^2}$. Figure~\ref{fig:unfolding} shows the case for two of those.

For the uncertainty estimate, the uncertainties of the $\theta_x^*$ and
$\theta_y^*$ resolutions (see Section~\ref{sec:resolution}) as well as
fit-model dependence have been taken into account. Altogether, the uncertainty
is smaller than $0.1\un{\%}$.


\subsubsection{Normalisation}
\label{sec:normalisation}

The normalisation factor $\mathcal{N}$ is determined by requiring the
integrated nuclear elastic cross-section to be $\sigma_{\rm el} = 31.0\un{mb}$
as obtained by TOTEM from a $\beta^* = 90\un{m}$ dataset at the same energy
\cite{totem-13tev-90m}. The elastic cross-section is extracted from the data in
two parts. The first part sums the $\d\sigma/\d t$ histogram bins for $0.01 <
|t| < 0.5\un{GeV^2}$. The second part corresponds to the integral over $0 < |t|
< 0.01\un{GeV^2}$ of an exponential fitted to the data on the interval $0.01 <
|t| < 0.05\un{GeV^2}$.

The uncertainty of $\mathcal{N}$ is dominated by the $5.5\un{\%}$ uncertainty
of $\sigma_{\rm el}$ from Ref.~\cite{totem-13tev-90m}.


\subsubsection{Binning}
\label{sec:binning}

The bin sizes are set according to the $t$ resolution. Three different binnings
are considered in this analysis: ``dense'' where the bin size is as large as
the standard deviation of $|t|$, ``medium'' with bins twice as large and
``coarse'' with bins three times larger than the standard deviation of $|t|$.


\subsection{Data Merging}
\label{sec:data merging}

\begin{figure*}
\vskip-5mm
\begin{center}
\includegraphics{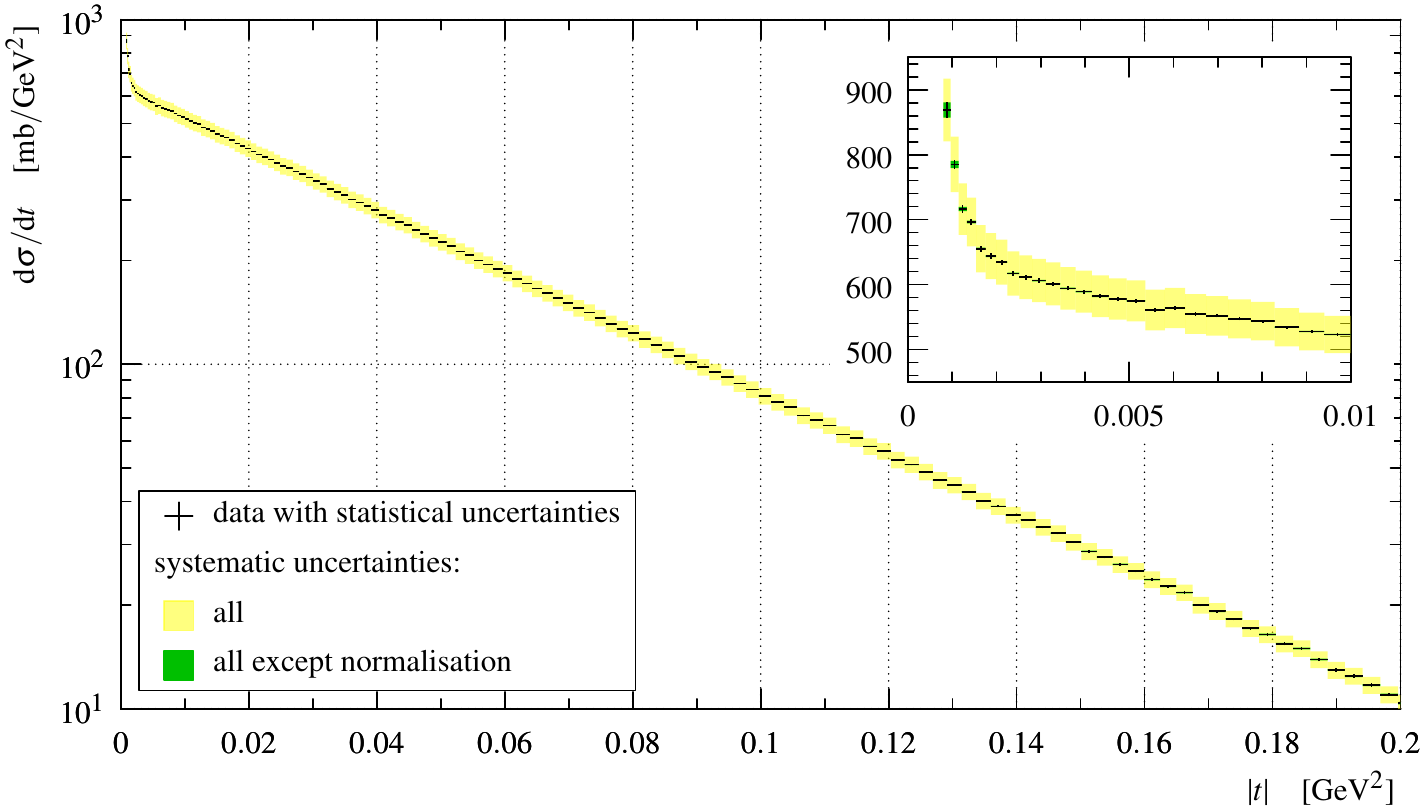}
\caption{%
Differential cross-section from Table \ref{tab:data} with statistical (bars)
	and systematic uncertainties (bands). The yellow band represents all
	systematic uncertainties, the green one all but normalisation. The
	bands are centred around the bin content. {\bf Inset}: a low-$|t|$ zoom
	of cross-section rise due to the Coulomb interaction.
	}
\label{fig:dsdt}
\end{center}
\end{figure*}

After analysing the data in each diagonal and LHC fill separately, the
individual differential cross-section distributions are merged. This is
accomplished by a per-bin weighted average, with the weight given by inverse
squared statistical uncertainty. The final cross-section values are listed in
Table~\ref{tab:data} and are visualised in Figure~\ref{fig:dsdt}. The figure
clearly shows a rapid cross-section rise below $|t| \lesssim 0.002\un{GeV^2}$
which, as interpreted later, is an effect due to the electromagnetic
interaction.


\subsection{Systematic Uncertainties}
\label{sec:systematics}

The following sources of systematic uncertainties have been considered.
\begin{itemize}[noitemsep,topsep=0pt]

\item Alignment: shifts in $\theta^*_{x,y}$ (see Section~\ref{sec:alignment}).
	Both left-right symmetric and anti-symmetric modes have been
		considered. In the vertical plane, both contributions
		correlated and uncorrelated between the diagonals have been
		considered.

\item Alignment $x$-$y$ tilts and optics: mixing between $\theta^*_{x}$ and
	$\theta^*_{y}$ (see Section~\ref{sec:alignment}). Both left-right
		symmetric and anti-symmetric modes have been considered.

\item Optics uncertainties: scaling of $\theta^*_{x,y}$ (see
	Section~\ref{sec:optics}). The three relevant modes in Eq.~(\ref{eq:opt
		bias}) have been considered.

\item Background subtraction (see Section~\ref{sec:background}): the
	$t$-dependent uncertainty of the correction factor $\mathcal{B}$.

\item Acceptance correction (see Section~\ref{sec:acc corr}): the uncertainty
	of resolution parameters, non-gaussianity of the resolution
		distributions, left-right asymmetry of the beam divergence.

\item Inefficiency corrections (see Section~\ref{sec:ineff corr}): for the
	uncorrelated inefficiency $\mathcal{I}_1$ both uncertainties of the
		fitted slope and intercept have been considered. For the
		correlated inefficiency $\mathcal{I}_2$ the uncertainty of its
		value has been considered.

\item The beam-momentum uncertainty: considered when the scattering angles are
	translated to $t$, see Eq.~(\ref{eq:th t}). The uncertainty was
		estimated by LHC experts as $0.1\un{\%}$ \cite{beam-mom-unc} in
		agreement with a previous assessment by TOTEM
		(Section~5.2.8.~in \cite{totem-8tev-90m}).

\item Unsmearing (see Section~\ref{sec:unfolding}): uncertainty of resolution
	parameters and model dependence of the fit.

\item Normalisation (see Section~\ref{sec:normalisation}): overall
	multiplicative factor.

\end{itemize}

For each error source, its effect on the $|t|$-distribution is evaluated with a
Monte-Carlo simulation. It uses a fit of the final differential cross-section
data to generate the true $t$-distribution and, in parallel, builds another
$t$-distribution where the systematic error at $1\un{\sigma}$ level is
introduced. The difference between the two $t$-distributions gives the
systematic effect on the differential cross-section. This procedure is formally
equivalent to evaluating
\begin{equation}
\label{eq:syst mode}
\delta s_{q}(t) \equiv \frac{\partial(\d\sigma/\d t)}{\partial q}\ \delta q\ ,
\end{equation}
where $\delta q$ corresponds to a $1\un{\sigma}$ bias in the quantity $q$
responsible for a given systematic effect.

The systematic uncertainty corresponding to the final differential
cross-section merged from all the analysed LHC fills and both diagonals is
propagated according to the same method as applied to the data, see
Section~\ref{sec:data merging}. To be conservative, the systematic errors are
assumed fully correlated among the four analysed LHC fills. The correlations
between the two diagonals are respected for each systematic effect. This is
particularly important for the vertical (mis)-alignment which is predominantly
anti-correlated between the diagonals. While this uncertainty in the lowest
$|t|$ bin reaches about $7\un{\%}$ for a single diagonal, once the diagonals
are merged the impact drops to about $1.2\un{\%}$.

\begin{figure*}
\begin{center}
\includegraphics{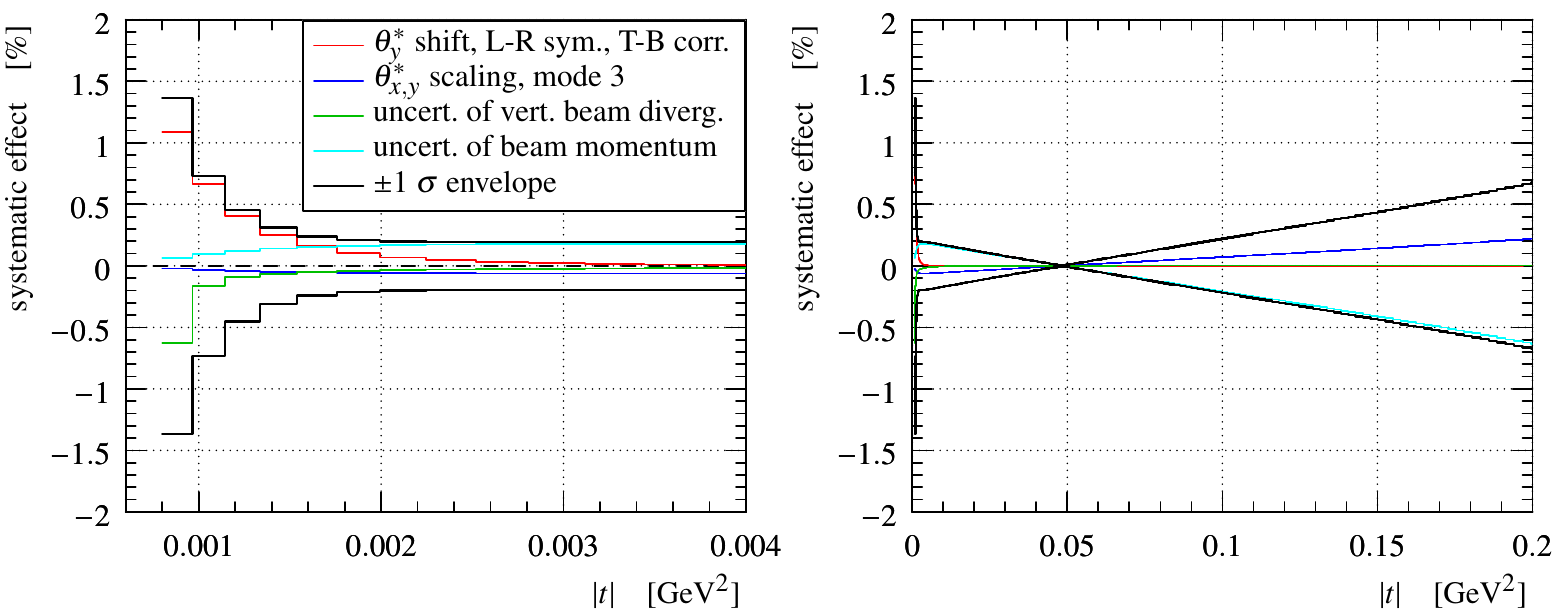}
\vskip-2mm
\caption{%
	Relative variation of the final differential cross-section due to
	systematic uncertainties (medium binning). The colourful histograms
	represent the leading uncertainties, each of them corresponds to a
	$1\un{\sigma}$ bias, cf.~Eq.~(\ref{eq:syst mode}). The envelope is
	determined by summing all considered contributions (except
	normalisation) in quadrature for each $|t|$ value.
	}
\label{fig:syst unc}
\end{center}
\end{figure*}

The leading uncertainties (except normalisation) are shown in
Figure~\ref{fig:syst unc}. At low $|t|$ they include the vertical alignment
(left-right symmetric, top-bottom correlated) and the uncertainty of the
vertical beam divergence. At higher $|t|$ values, the uncertainties are
dominated by the beam momentum and optics uncertainties (mode 3 in
Eq.~(\ref{eq:opt bias modes})). These leading effects are listed in
Table~\ref{tab:data} which can be used to approximate the covariance matrix of
systematic uncertainties:
\begin{equation}
\label{eq:covar mat}
\mat V_{ij} = \sum_{q} \delta s_{q}(i)\ \delta s_{q}(j)\: ,
\end{equation}
where $i$ and $j$ are bin indices (row numbers in Table~\ref{tab:data}) and the
sum goes over the leading error contributions $q$ (five rightmost columns in
the table).

\subsection{Systematic Cross-Checks}
\label{sec:cross checks}

\begin{figure*}
\vskip-4mm
\begin{center}
\includegraphics{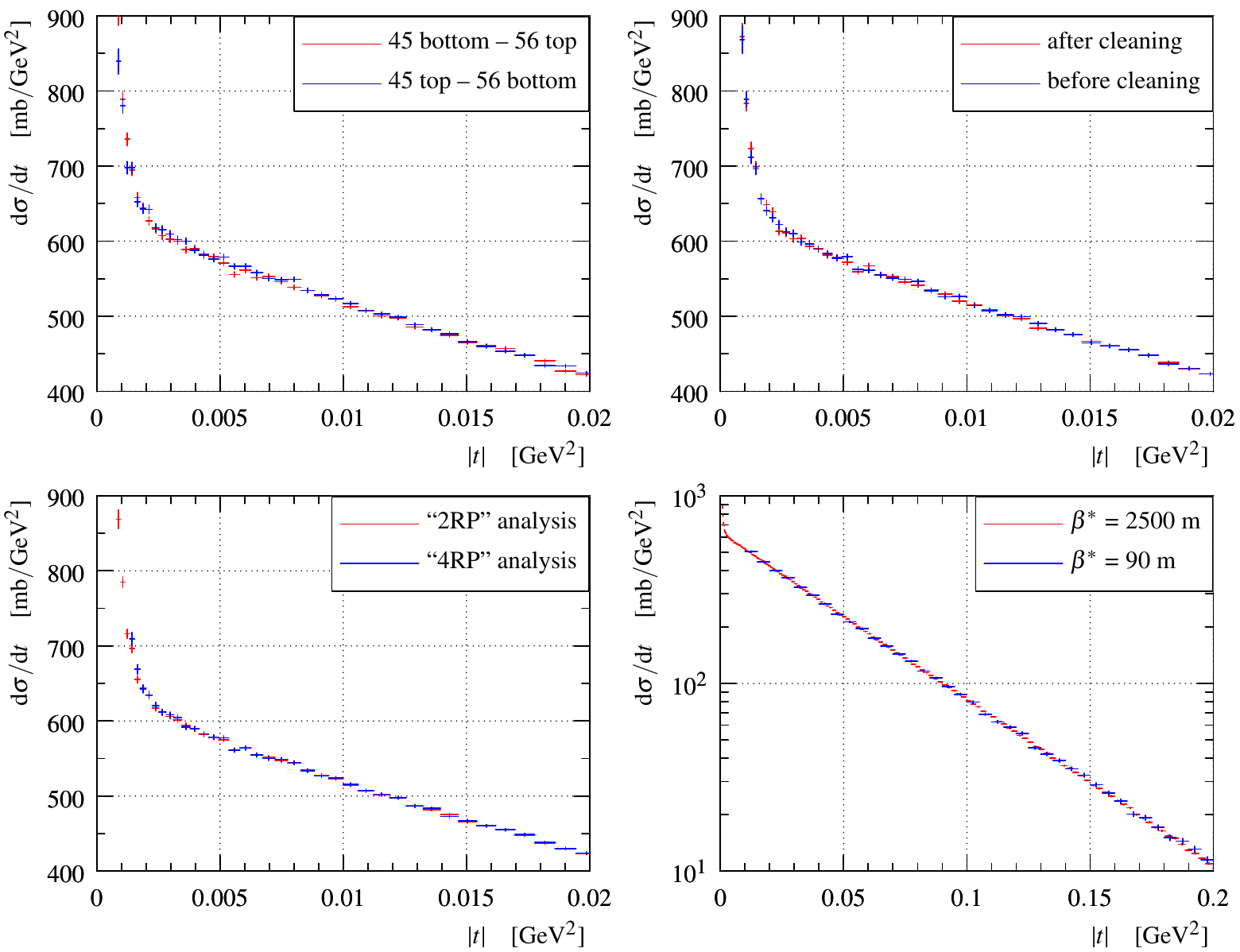}
\vskip-2mm
\caption{%
	Collection of cross-check plots (medium binning, only statistical
	uncertainties are plotted).  {\bf Top left}: comparison of results from
	the two diagonals, data from all LHC fills.  {\bf Top right}:
	comparison of results from time periods after and before beam
	cleanings, data from all LHC fills and both diagonals.  {\bf Bottom
	left}: comparison of results from ``2RP'' and ``4RP'' analyses, data
	from all LHC fills and both diagonals.  {\bf Bottom right}: comparison
	of results obtained from two different data-takings at the same energy
	but with different optics. The blue histogram is taken from
	Ref.~\cite{totem-13tev-90m}.
	}
\label{fig:dsdt checks}
\end{center}
\end{figure*}

Compatible results have been obtained by analysing data subsets of events from
different bunches, different diagonals (Figure~\ref{fig:dsdt checks}, top
left), different fills and different time periods -- in particular those right
after and right before the beam cleanings (Figure~\ref{fig:dsdt checks}, top
right). Figure~\ref{fig:dsdt checks}, bottom left, shows that both analysis
approaches, ``2RP'' and ``4RP'', yield compatible results. The relatively large
difference between the diagonals at very low $|t|$ (Figure~\ref{fig:dsdt
checks}, top left) is fully within the uncertainty due to the vertical
misalignment, see Section~\ref{sec:systematics}.

Figure~\ref{fig:dsdt checks}, bottom right, shows an excellent agreement
between the data from this analysis and previous results obtained with $\beta^*
= 90\un{m}$ optics \cite{totem-13tev-90m}.

\def\Header
{
\footnotesize
\setlength{\tabcolsep}{4.5pt}
\def\arraystretch{0.8}
\begin{tabular}{ccc@{\hskip15pt}cccccccc}
\hline
	\multispan3\vrule width0pt height9pt depth3pt\hss $|t|$ bin $\unt{GeV^2}$\hss & \multispan8\hss $\d\sigma/\d t \ung{mb/GeV^2}$ \hss \cr
	\multispan3\hrulefill\hbox to12pt{\hfil} & \multispan8\hrulefill\cr
	left & right & represent. & value & statist.     & system.  & normal.    & alignment	& optics	& vert.~beam	& beam	\cr
	edge & edge  & point      &       & uncert.      & uncert.  &            & vert.~shift	& mode 3	& divergence	& mom.	\cr
\hline
}

\begin{table*}
\caption{%
	The elastic differential cross-section as determined in this analysis
	(medium binning). The three leftmost columns describe the bins in $t$.
	The representative point gives the $t$ value suitable for
	fitting~\cite{lafferty94}. The other columns are related to the
	differential cross-section. The five rightmost columns give the leading
	systematic biases in $\d\sigma/\d t$ for $1\sigma$-shifts in the
	respective quantities, $\delta s_q$, see Eqs.~(\ref{eq:syst mode}) and
	(\ref{eq:covar mat}). The contribution due to optics corresponds to the
	third vector in Eq.~(\ref{eq:opt bias modes}). In order to avoid
	undesired interplay between statistical and systematic uncertainties,
	the latter are calculated from the relative uncertainties
	(Section~\ref{sec:systematics}) by multiplying by a smooth fit
	(Figure~\ref{fig:fit exp3 0.15}) evaluated at the representative point.
	}%
\vskip-3mm
\label{tab:data}
\begin{center}
\Header
\vrule width0pt height8pt
$0.000800$ & $0.000966$ & $0.000879$ & $868.726$ & $12.518$ & $48.472$ & $+46.865$ & $+9.265$ & $-0.175$ & $-5.360$ & $+0.548$ \cr
$0.000966$ & $0.001144$ & $0.001051$ & $784.894$ & $7.252$ & $42.786$ & $+42.318$ & $+5.098$ & $-0.252$ & $-1.279$ & $+0.750$ \cr
$0.001144$ & $0.001335$ & $0.001236$ & $716.217$ & $5.943$ & $39.656$ & $+39.476$ & $+2.900$ & $-0.299$ & $-0.660$ & $+0.876$ \cr
$0.001335$ & $0.001540$ & $0.001434$ & $696.283$ & $5.279$ & $37.685$ & $+37.603$ & $+1.722$ & $-0.330$ & $-0.435$ & $+0.963$ \cr
$0.001540$ & $0.001759$ & $0.001646$ & $655.272$ & $4.710$ & $36.358$ & $+36.313$ & $+1.059$ & $-0.350$ & $-0.327$ & $+1.012$ \cr
$0.001759$ & $0.001995$ & $0.001874$ & $643.657$ & $4.346$ & $35.415$ & $+35.385$ & $+0.670$ & $-0.363$ & $-0.259$ & $+1.047$ \cr
$0.001995$ & $0.002248$ & $0.002118$ & $634.502$ & $4.047$ & $34.713$ & $+34.689$ & $+0.435$ & $-0.370$ & $-0.212$ & $+1.068$ \cr
$0.002248$ & $0.002519$ & $0.002380$ & $617.090$ & $3.764$ & $34.166$ & $+34.144$ & $+0.287$ & $-0.375$ & $-0.180$ & $+1.080$ \cr
$0.002519$ & $0.002809$ & $0.002661$ & $611.317$ & $3.552$ & $33.720$ & $+33.699$ & $+0.193$ & $-0.377$ & $-0.156$ & $+1.085$ \cr
$0.002809$ & $0.003117$ & $0.002960$ & $606.121$ & $3.374$ & $33.341$ & $+33.320$ & $+0.132$ & $-0.377$ & $-0.137$ & $+1.085$ \cr
$0.003117$ & $0.003444$ & $0.003279$ & $601.057$ & $3.212$ & $33.005$ & $+32.984$ & $+0.092$ & $-0.375$ & $-0.122$ & $+1.080$ \cr
$0.003444$ & $0.003791$ & $0.003616$ & $594.143$ & $3.064$ & $32.695$ & $+32.675$ & $+0.065$ & $-0.373$ & $-0.109$ & $+1.073$ \cr
$0.003791$ & $0.004155$ & $0.003972$ & $589.140$ & $2.945$ & $32.402$ & $+32.382$ & $+0.047$ & $-0.369$ & $-0.099$ & $+1.062$ \cr
$0.004155$ & $0.004538$ & $0.004346$ & $581.891$ & $2.827$ & $32.117$ & $+32.097$ & $+0.033$ & $-0.365$ & $-0.090$ & $+1.050$ \cr
$0.004538$ & $0.004940$ & $0.004738$ & $577.737$ & $2.726$ & $31.836$ & $+31.816$ & $+0.024$ & $-0.360$ & $-0.082$ & $+1.035$ \cr
$0.004940$ & $0.005361$ & $0.005150$ & $575.008$ & $2.636$ & $31.553$ & $+31.534$ & $+0.019$ & $-0.354$ & $-0.075$ & $+1.019$ \cr
$0.005361$ & $0.005801$ & $0.005581$ & $560.883$ & $2.526$ & $31.266$ & $+31.248$ & $+0.016$ & $-0.349$ & $-0.066$ & $+1.002$ \cr
$0.005801$ & $0.006260$ & $0.006030$ & $563.968$ & $2.468$ & $30.974$ & $+30.956$ & $+0.014$ & $-0.342$ & $-0.059$ & $+0.984$ \cr
$0.006260$ & $0.006737$ & $0.006498$ & $554.645$ & $2.387$ & $30.676$ & $+30.659$ & $+0.012$ & $-0.335$ & $-0.053$ & $+0.965$ \cr
$0.006737$ & $0.007232$ & $0.006984$ & $551.682$ & $2.323$ & $30.372$ & $+30.355$ & $+0.011$ & $-0.329$ & $-0.048$ & $+0.945$ \cr
$0.007232$ & $0.007746$ & $0.007488$ & $547.232$ & $2.260$ & $30.060$ & $+30.043$ & $+0.010$ & $-0.321$ & $-0.042$ & $+0.924$ \cr
$0.007746$ & $0.008279$ & $0.008012$ & $543.798$ & $2.202$ & $29.739$ & $+29.723$ & $+0.009$ & $-0.314$ & $-0.036$ & $+0.903$ \cr
$0.008279$ & $0.008833$ & $0.008556$ & $534.391$ & $2.133$ & $29.410$ & $+29.395$ & $+0.008$ & $-0.306$ & $-0.032$ & $+0.881$ \cr
$0.008833$ & $0.009407$ & $0.009120$ & $527.706$ & $2.076$ & $29.073$ & $+29.059$ & $+0.008$ & $-0.299$ & $-0.029$ & $+0.859$ \cr
$0.009407$ & $0.009999$ & $0.009703$ & $523.040$ & $2.027$ & $28.729$ & $+28.715$ & $+0.007$ & $-0.291$ & $-0.026$ & $+0.836$ \cr
$0.009999$ & $0.010608$ & $0.010303$ & $514.667$ & $1.976$ & $28.377$ & $+28.364$ & $+0.006$ & $-0.283$ & $-0.023$ & $+0.813$ \cr
$0.010608$ & $0.011237$ & $0.010922$ & $507.673$ & $1.925$ & $28.019$ & $+28.006$ & $+0.005$ & $-0.274$ & $-0.020$ & $+0.789$ \cr
$0.011237$ & $0.011887$ & $0.011562$ & $501.645$ & $1.877$ & $27.653$ & $+27.641$ & $+0.005$ & $-0.266$ & $-0.018$ & $+0.765$ \cr
$0.011887$ & $0.012556$ & $0.012221$ & $498.095$ & $1.840$ & $27.280$ & $+27.269$ & $+0.004$ & $-0.258$ & $-0.015$ & $+0.741$ \cr
$0.012556$ & $0.013242$ & $0.012898$ & $487.164$ & $1.791$ & $26.902$ & $+26.891$ & $+0.003$ & $-0.249$ & $-0.013$ & $+0.717$ \cr
$0.013242$ & $0.013948$ & $0.013594$ & $482.155$ & $1.753$ & $26.519$ & $+26.509$ & $+0.003$ & $-0.241$ & $-0.011$ & $+0.692$ \cr
$0.013948$ & $0.014674$ & $0.014311$ & $475.608$ & $1.712$ & $26.130$ & $+26.120$ & $+0.003$ & $-0.233$ & $-0.009$ & $+0.668$ \cr
$0.014674$ & $0.015421$ & $0.015047$ & $465.619$ & $1.668$ & $25.735$ & $+25.726$ & $+0.002$ & $-0.224$ & $-0.007$ & $+0.643$ \cr
$0.015421$ & $0.016186$ & $0.015803$ & $460.386$ & $1.635$ & $25.335$ & $+25.327$ & $+0.002$ & $-0.215$ & $-0.005$ & $+0.618$ \cr
$0.016186$ & $0.016969$ & $0.016577$ & $455.279$ & $1.605$ & $24.933$ & $+24.925$ & $+0.001$ & $-0.207$ & $-0.004$ & $+0.594$ \cr
$0.016969$ & $0.017771$ & $0.017370$ & $447.960$ & $1.570$ & $24.527$ & $+24.519$ & $+0.001$ & $-0.198$ & $-0.003$ & $+0.569$ \cr
$0.017771$ & $0.018597$ & $0.018183$ & $437.466$ & $1.526$ & $24.116$ & $+24.109$ & $+0.001$ & $-0.190$ & $-0.002$ & $+0.545$ \cr
$0.018597$ & $0.019443$ & $0.019020$ & $430.342$ & $1.493$ & $23.702$ & $+23.695$ & $+0.000$ & $-0.181$ & $-0.002$ & $+0.521$ \cr
$0.019443$ & $0.020308$ & $0.019874$ & $423.167$ & $1.463$ & $23.285$ & $+23.279$ & $+0.000$ & $-0.173$ & $-0.001$ & $+0.496$ \cr
$0.020308$ & $0.021189$ & $0.020748$ & $414.878$ & $1.432$ & $22.868$ & $+22.862$ & $-0.000$ & $-0.164$ & $-0.001$ & $+0.472$ \cr
$0.021189$ & $0.022087$ & $0.021638$ & $406.158$ & $1.402$ & $22.450$ & $+22.444$ & $-0.001$ & $-0.156$ & $-0.000$ & $+0.448$ \cr
$0.022087$ & $0.023007$ & $0.022547$ & $400.652$ & $1.374$ & $22.030$ & $+22.025$ & $-0.001$ & $-0.148$ & $+0.000$ & $+0.425$ \cr
$0.023007$ & $0.023942$ & $0.023475$ & $393.124$ & $1.348$ & $21.610$ & $+21.605$ & $-0.001$ & $-0.140$ & $+0.001$ & $+0.402$ \cr
$0.023942$ & $0.024899$ & $0.024421$ & $384.736$ & $1.317$ & $21.189$ & $+21.186$ & $-0.001$ & $-0.132$ & $+0.001$ & $+0.379$ \cr
$0.024899$ & $0.025878$ & $0.025389$ & $376.243$ & $1.286$ & $20.768$ & $+20.764$ & $-0.001$ & $-0.124$ & $+0.001$ & $+0.356$ \cr
$0.025878$ & $0.026875$ & $0.026377$ & $372.189$ & $1.266$ & $20.346$ & $+20.343$ & $-0.002$ & $-0.116$ & $+0.002$ & $+0.334$ \cr
$0.026875$ & $0.027895$ & $0.027385$ & $362.930$ & $1.235$ & $19.925$ & $+19.922$ & $-0.002$ & $-0.108$ & $+0.002$ & $+0.312$ \cr
$0.027895$ & $0.028932$ & $0.028413$ & $357.126$ & $1.214$ & $19.505$ & $+19.502$ & $-0.002$ & $-0.101$ & $+0.002$ & $+0.290$ \cr
$0.028932$ & $0.029988$ & $0.029460$ & $348.345$ & $1.186$ & $19.086$ & $+19.084$ & $-0.002$ & $-0.094$ & $+0.003$ & $+0.269$ \cr
$0.029988$ & $0.031067$ & $0.030528$ & $339.830$ & $1.158$ & $18.668$ & $+18.666$ & $-0.002$ & $-0.086$ & $+0.003$ & $+0.248$ \cr
$0.031067$ & $0.032162$ & $0.031615$ & $333.025$ & $1.137$ & $18.252$ & $+18.250$ & $-0.002$ & $-0.079$ & $+0.003$ & $+0.228$ \cr
$0.032162$ & $0.033279$ & $0.032720$ & $323.442$ & $1.109$ & $17.838$ & $+17.837$ & $-0.002$ & $-0.072$ & $+0.003$ & $+0.208$ \cr
$0.033279$ & $0.034415$ & $0.033846$ & $316.769$ & $1.087$ & $17.427$ & $+17.426$ & $-0.002$ & $-0.066$ & $+0.003$ & $+0.189$ \cr
$0.034415$ & $0.035568$ & $0.034989$ & $309.514$ & $1.066$ & $17.019$ & $+17.018$ & $-0.002$ & $-0.059$ & $+0.003$ & $+0.170$ \cr
$0.035568$ & $0.036742$ & $0.036154$ & $300.609$ & $1.040$ & $16.614$ & $+16.613$ & $-0.002$ & $-0.053$ & $+0.003$ & $+0.151$ \cr
$0.036742$ & $0.037930$ & $0.037335$ & $295.114$ & $1.024$ & $16.213$ & $+16.213$ & $-0.002$ & $-0.047$ & $+0.003$ & $+0.133$ \cr
$0.037930$ & $0.039138$ & $0.038533$ & $288.375$ & $1.003$ & $15.817$ & $+15.816$ & $-0.003$ & $-0.041$ & $+0.004$ & $+0.116$ \cr
$0.039138$ & $0.040369$ & $0.039752$ & $280.807$ & $0.979$ & $15.423$ & $+15.423$ & $-0.003$ & $-0.035$ & $+0.004$ & $+0.099$ \cr
$0.040369$ & $0.041618$ & $0.040990$ & $271.508$ & $0.955$ & $15.033$ & $+15.033$ & $-0.003$ & $-0.029$ & $+0.004$ & $+0.083$ \cr
$0.041618$ & $0.042887$ & $0.042251$ & $266.133$ & $0.938$ & $14.647$ & $+14.647$ & $-0.003$ & $-0.024$ & $+0.004$ & $+0.067$ \cr
$0.042887$ & $0.044177$ & $0.043531$ & $258.862$ & $0.917$ & $14.266$ & $+14.266$ & $-0.003$ & $-0.018$ & $+0.004$ & $+0.052$ \cr
$0.044177$ & $0.045487$ & $0.044830$ & $253.719$ & $0.900$ & $13.889$ & $+13.888$ & $-0.003$ & $-0.013$ & $+0.004$ & $+0.038$ \cr
$0.045487$ & $0.046815$ & $0.046149$ & $245.394$ & $0.879$ & $13.516$ & $+13.516$ & $-0.003$ & $-0.008$ & $+0.004$ & $+0.024$ \cr
$0.046815$ & $0.048165$ & $0.047489$ & $238.906$ & $0.860$ & $13.148$ & $+13.148$ & $-0.003$ & $-0.004$ & $+0.004$ & $+0.010$ \cr
$0.048165$ & $0.049528$ & $0.048844$ & $232.195$ & $0.843$ & $12.786$ & $+12.786$ & $-0.003$ & $+0.001$ & $+0.004$ & $-0.003$ \cr
$0.049528$ & $0.050917$ & $0.050221$ & $226.191$ & $0.824$ & $12.430$ & $+12.430$ & $-0.003$ & $+0.005$ & $+0.004$ & $-0.015$ \cr
$0.050917$ & $0.052322$ & $0.051619$ & $220.655$ & $0.809$ & $12.078$ & $+12.078$ & $-0.002$ & $+0.009$ & $+0.004$ & $-0.027$ \cr
\hline
\end{tabular}
\end{center}
\end{table*}

\begin{table*}
\caption{%
Continuation of Table~\ref{tab:data}.
}%
\vskip-3mm
\label{tab:data cont}
\begin{center}
\Header
\vrule width0pt height8pt
$0.052322$ & $0.053748$ & $0.053031$ & $212.493$ & $0.787$ & $11.732$ & $+11.732$ & $-0.002$ & $+0.013$ & $+0.003$ & $-0.039$ \cr
$0.053748$ & $0.055193$ & $0.054470$ & $207.171$ & $0.772$ & $11.391$ & $+11.391$ & $-0.002$ & $+0.017$ & $+0.003$ & $-0.049$ \cr
$0.055193$ & $0.056660$ & $0.055923$ & $200.154$ & $0.752$ & $11.056$ & $+11.056$ & $-0.002$ & $+0.021$ & $+0.003$ & $-0.060$ \cr
$0.056660$ & $0.058145$ & $0.057401$ & $194.826$ & $0.737$ & $10.726$ & $+10.726$ & $-0.002$ & $+0.024$ & $+0.003$ & $-0.069$ \cr
$0.058145$ & $0.059649$ & $0.058894$ & $189.250$ & $0.722$ & $10.403$ & $+10.402$ & $-0.002$ & $+0.027$ & $+0.003$ & $-0.079$ \cr
$0.059649$ & $0.061175$ & $0.060411$ & $184.095$ & $0.706$ & $10.085$ & $+10.084$ & $-0.002$ & $+0.030$ & $+0.003$ & $-0.087$ \cr
$0.061175$ & $0.062717$ & $0.061942$ & $177.115$ & $0.689$ & $9.773$ & $+9.773$ & $-0.002$ & $+0.033$ & $+0.003$ & $-0.096$ \cr
$0.062717$ & $0.064277$ & $0.063496$ & $171.504$ & $0.674$ & $9.468$ & $+9.467$ & $-0.002$ & $+0.036$ & $+0.003$ & $-0.103$ \cr
$0.064277$ & $0.065859$ & $0.065065$ & $165.886$ & $0.658$ & $9.169$ & $+9.168$ & $-0.002$ & $+0.038$ & $+0.003$ & $-0.110$ \cr
$0.065859$ & $0.067461$ & $0.066659$ & $160.981$ & $0.644$ & $8.875$ & $+8.874$ & $-0.002$ & $+0.041$ & $+0.003$ & $-0.117$ \cr
$0.067461$ & $0.069082$ & $0.068270$ & $155.821$ & $0.629$ & $8.588$ & $+8.587$ & $-0.002$ & $+0.043$ & $+0.003$ & $-0.123$ \cr
$0.069082$ & $0.070723$ & $0.069900$ & $150.892$ & $0.615$ & $8.307$ & $+8.306$ & $-0.002$ & $+0.045$ & $+0.003$ & $-0.129$ \cr
$0.070723$ & $0.072392$ & $0.071556$ & $145.575$ & $0.599$ & $8.031$ & $+8.030$ & $-0.002$ & $+0.047$ & $+0.003$ & $-0.134$ \cr
$0.072392$ & $0.074077$ & $0.073232$ & $141.394$ & $0.587$ & $7.762$ & $+7.760$ & $-0.002$ & $+0.048$ & $+0.003$ & $-0.139$ \cr
$0.074077$ & $0.075777$ & $0.074923$ & $136.424$ & $0.574$ & $7.499$ & $+7.497$ & $-0.002$ & $+0.050$ & $+0.003$ & $-0.144$ \cr
$0.075777$ & $0.077497$ & $0.076635$ & $131.196$ & $0.560$ & $7.242$ & $+7.241$ & $-0.002$ & $+0.051$ & $+0.003$ & $-0.148$ \cr
$0.077497$ & $0.079239$ & $0.078366$ & $126.732$ & $0.546$ & $6.992$ & $+6.990$ & $-0.002$ & $+0.053$ & $+0.003$ & $-0.151$ \cr
$0.079239$ & $0.080997$ & $0.080115$ & $123.202$ & $0.536$ & $6.747$ & $+6.745$ & $-0.001$ & $+0.054$ & $+0.003$ & $-0.155$ \cr
$0.080997$ & $0.082783$ & $0.081887$ & $118.162$ & $0.521$ & $6.509$ & $+6.507$ & $-0.001$ & $+0.055$ & $+0.003$ & $-0.157$ \cr
$0.082783$ & $0.084581$ & $0.083680$ & $114.017$ & $0.510$ & $6.276$ & $+6.274$ & $-0.001$ & $+0.056$ & $+0.003$ & $-0.160$ \cr
$0.084581$ & $0.086404$ & $0.085491$ & $110.242$ & $0.497$ & $6.050$ & $+6.047$ & $-0.001$ & $+0.056$ & $+0.002$ & $-0.162$ \cr
$0.086404$ & $0.088239$ & $0.087319$ & $105.794$ & $0.486$ & $5.830$ & $+5.827$ & $-0.001$ & $+0.057$ & $+0.002$ & $-0.164$ \cr
$0.088239$ & $0.090099$ & $0.089166$ & $101.754$ & $0.473$ & $5.615$ & $+5.613$ & $-0.001$ & $+0.057$ & $+0.002$ & $-0.165$ \cr
$0.090099$ & $0.091976$ & $0.091033$ & $98.155$ & $0.462$ & $5.407$ & $+5.404$ & $-0.001$ & $+0.058$ & $+0.002$ & $-0.166$ \cr
$0.091976$ & $0.093868$ & $0.092918$ & $95.128$ & $0.453$ & $5.204$ & $+5.201$ & $-0.001$ & $+0.058$ & $+0.002$ & $-0.167$ \cr
$0.093868$ & $0.095784$ & $0.094823$ & $91.683$ & $0.442$ & $5.008$ & $+5.005$ & $-0.001$ & $+0.058$ & $+0.002$ & $-0.168$ \cr
$0.095784$ & $0.097721$ & $0.096750$ & $87.967$ & $0.430$ & $4.816$ & $+4.813$ & $-0.001$ & $+0.058$ & $+0.002$ & $-0.168$ \cr
$0.097721$ & $0.099679$ & $0.098697$ & $84.557$ & $0.420$ & $4.630$ & $+4.627$ & $-0.001$ & $+0.058$ & $+0.002$ & $-0.168$ \cr
$0.099679$ & $0.101659$ & $0.100666$ & $80.860$ & $0.408$ & $4.450$ & $+4.446$ & $-0.001$ & $+0.058$ & $+0.002$ & $-0.168$ \cr
$0.101659$ & $0.103658$ & $0.102656$ & $77.695$ & $0.398$ & $4.275$ & $+4.271$ & $-0.001$ & $+0.058$ & $+0.002$ & $-0.168$ \cr
$0.103658$ & $0.105679$ & $0.104666$ & $75.105$ & $0.389$ & $4.105$ & $+4.101$ & $-0.001$ & $+0.058$ & $+0.002$ & $-0.167$ \cr
$0.105679$ & $0.107705$ & $0.106689$ & $71.021$ & $0.378$ & $3.940$ & $+3.936$ & $-0.001$ & $+0.058$ & $+0.002$ & $-0.166$ \cr
$0.107705$ & $0.109766$ & $0.108732$ & $68.777$ & $0.368$ & $3.781$ & $+3.777$ & $-0.001$ & $+0.057$ & $+0.002$ & $-0.165$ \cr
$0.109766$ & $0.111845$ & $0.110802$ & $66.428$ & $0.360$ & $3.627$ & $+3.622$ & $-0.001$ & $+0.057$ & $+0.002$ & $-0.164$ \cr
$0.111845$ & $0.113945$ & $0.112891$ & $62.555$ & $0.348$ & $3.477$ & $+3.473$ & $-0.001$ & $+0.056$ & $+0.002$ & $-0.162$ \cr
$0.113945$ & $0.116056$ & $0.114995$ & $60.954$ & $0.342$ & $3.332$ & $+3.328$ & $-0.000$ & $+0.056$ & $+0.002$ & $-0.161$ \cr
$0.116056$ & $0.118185$ & $0.117116$ & $57.750$ & $0.332$ & $3.193$ & $+3.189$ & $-0.000$ & $+0.055$ & $+0.002$ & $-0.159$ \cr
$0.118185$ & $0.120342$ & $0.119260$ & $55.934$ & $0.324$ & $3.058$ & $+3.054$ & $-0.000$ & $+0.055$ & $+0.001$ & $-0.157$ \cr
$0.120342$ & $0.122517$ & $0.121427$ & $52.788$ & $0.314$ & $2.928$ & $+2.923$ & $-0.000$ & $+0.054$ & $+0.001$ & $-0.155$ \cr
$0.122517$ & $0.124719$ & $0.123615$ & $51.158$ & $0.307$ & $2.801$ & $+2.797$ & $-0.000$ & $+0.053$ & $+0.001$ & $-0.153$ \cr
$0.124719$ & $0.126932$ & $0.125821$ & $48.734$ & $0.299$ & $2.680$ & $+2.675$ & $-0.000$ & $+0.052$ & $+0.001$ & $-0.150$ \cr
$0.126932$ & $0.129175$ & $0.128048$ & $46.061$ & $0.288$ & $2.562$ & $+2.557$ & $-0.000$ & $+0.051$ & $+0.001$ & $-0.148$ \cr
$0.129175$ & $0.131421$ & $0.130294$ & $44.625$ & $0.284$ & $2.449$ & $+2.444$ & $-0.000$ & $+0.051$ & $+0.001$ & $-0.146$ \cr
$0.131421$ & $0.133681$ & $0.132548$ & $42.645$ & $0.276$ & $2.340$ & $+2.335$ & $-0.000$ & $+0.050$ & $+0.001$ & $-0.143$ \cr
$0.133681$ & $0.135974$ & $0.134823$ & $40.137$ & $0.266$ & $2.236$ & $+2.230$ & $-0.000$ & $+0.049$ & $+0.001$ & $-0.140$ \cr
$0.135974$ & $0.138285$ & $0.137125$ & $38.674$ & $0.260$ & $2.134$ & $+2.129$ & $-0.000$ & $+0.048$ & $+0.001$ & $-0.138$ \cr
$0.138285$ & $0.140614$ & $0.139446$ & $36.488$ & $0.252$ & $2.037$ & $+2.032$ & $-0.000$ & $+0.047$ & $+0.001$ & $-0.135$ \cr
$0.140614$ & $0.142962$ & $0.141784$ & $35.400$ & $0.247$ & $1.943$ & $+1.938$ & $-0.000$ & $+0.046$ & $+0.001$ & $-0.132$ \cr
$0.142962$ & $0.145328$ & $0.144140$ & $33.650$ & $0.240$ & $1.853$ & $+1.848$ & $-0.000$ & $+0.045$ & $+0.001$ & $-0.129$ \cr
$0.145328$ & $0.147710$ & $0.146515$ & $32.302$ & $0.234$ & $1.766$ & $+1.761$ & $-0.000$ & $+0.044$ & $+0.001$ & $-0.126$ \cr
$0.147710$ & $0.150118$ & $0.148909$ & $30.473$ & $0.226$ & $1.683$ & $+1.678$ & $+0.000$ & $+0.043$ & $+0.001$ & $-0.123$ \cr
$0.150118$ & $0.152551$ & $0.151330$ & $28.634$ & $0.218$ & $1.603$ & $+1.597$ & $+0.000$ & $+0.042$ & $+0.001$ & $-0.121$ \cr
$0.152551$ & $0.155000$ & $0.153771$ & $27.551$ & $0.213$ & $1.525$ & $+1.520$ & $+0.000$ & $+0.041$ & $+0.001$ & $-0.118$ \cr
$0.155000$ & $0.157452$ & $0.156221$ & $26.250$ & $0.208$ & $1.451$ & $+1.446$ & $+0.000$ & $+0.040$ & $+0.001$ & $-0.115$ \cr
$0.157452$ & $0.159942$ & $0.158693$ & $25.092$ & $0.202$ & $1.380$ & $+1.375$ & $+0.000$ & $+0.039$ & $+0.001$ & $-0.112$ \cr
$0.159942$ & $0.162445$ & $0.161189$ & $23.721$ & $0.195$ & $1.312$ & $+1.307$ & $+0.000$ & $+0.038$ & $+0.001$ & $-0.109$ \cr
$0.162445$ & $0.164974$ & $0.163705$ & $22.677$ & $0.190$ & $1.246$ & $+1.241$ & $+0.000$ & $+0.037$ & $+0.001$ & $-0.106$ \cr
$0.164974$ & $0.167515$ & $0.166239$ & $21.752$ & $0.186$ & $1.183$ & $+1.178$ & $+0.000$ & $+0.036$ & $+0.001$ & $-0.103$ \cr
$0.167515$ & $0.170078$ & $0.168791$ & $20.011$ & $0.177$ & $1.123$ & $+1.118$ & $+0.000$ & $+0.035$ & $+0.001$ & $-0.100$ \cr
$0.170078$ & $0.172669$ & $0.171369$ & $19.180$ & $0.173$ & $1.065$ & $+1.060$ & $+0.000$ & $+0.034$ & $+0.001$ & $-0.097$ \cr
$0.172669$ & $0.175277$ & $0.173966$ & $18.237$ & $0.168$ & $1.010$ & $+1.005$ & $+0.000$ & $+0.033$ & $+0.001$ & $-0.094$ \cr
$0.175277$ & $0.177899$ & $0.176583$ & $17.122$ & $0.162$ & $0.957$ & $+0.952$ & $+0.000$ & $+0.032$ & $+0.001$ & $-0.091$ \cr
$0.177899$ & $0.180548$ & $0.179217$ & $16.427$ & $0.158$ & $0.906$ & $+0.901$ & $+0.000$ & $+0.031$ & $+0.001$ & $-0.088$ \cr
$0.180548$ & $0.183212$ & $0.181874$ & $15.434$ & $0.153$ & $0.858$ & $+0.853$ & $+0.000$ & $+0.030$ & $+0.000$ & $-0.085$ \cr
$0.183212$ & $0.185903$ & $0.184552$ & $14.965$ & $0.149$ & $0.812$ & $+0.807$ & $+0.000$ & $+0.029$ & $+0.000$ & $-0.082$ \cr
$0.185903$ & $0.188606$ & $0.187249$ & $13.905$ & $0.144$ & $0.768$ & $+0.763$ & $+0.000$ & $+0.028$ & $+0.000$ & $-0.079$ \cr
$0.188606$ & $0.191329$ & $0.189960$ & $12.957$ & $0.138$ & $0.726$ & $+0.721$ & $+0.000$ & $+0.027$ & $+0.000$ & $-0.077$ \cr
$0.191329$ & $0.194088$ & $0.192702$ & $12.445$ & $0.135$ & $0.685$ & $+0.681$ & $+0.000$ & $+0.026$ & $+0.000$ & $-0.074$ \cr
$0.194088$ & $0.196855$ & $0.195466$ & $11.711$ & $0.130$ & $0.647$ & $+0.643$ & $+0.000$ & $+0.025$ & $+0.000$ & $-0.071$ \cr
$0.196855$ & $0.199646$ & $0.198244$ & $10.987$ & $0.126$ & $0.611$ & $+0.606$ & $+0.000$ & $+0.024$ & $+0.000$ & $-0.069$ \cr
$0.199646$ & $0.202452$ & $0.201041$ & $10.371$ & $0.122$ & $0.576$ & $+0.572$ & $+0.000$ & $+0.023$ & $+0.000$ & $-0.066$ \cr
\hline
\end{tabular}
\end{center}
\end{table*}

\section{Determination of $\rho$ and total cross-section}
\label{sec:rho}

The value of the $\rho$ parameter can be extracted from the differential
cross-section thanks to the effects of Coulomb-nuclear interference (CNI).
Explicit treatment of these effects allows also for a conceptually more
accurate determination of the total cross-section. 

Our modelling of the CNI effects is summarised in Section \ref{sec:rho cni},
Sections~\ref{sec:rho anal} and \ref{sec:rho anal norm var} describe data fits
and results. In Section~\ref{sec:rho anal} the differential cross-section
normalisation is fixed by the $\beta^* = 90\un{m}$ data \cite{totem-13tev-90m}
(see Section~\ref{sec:normalisation}). In Section~\ref{sec:rho anal norm var}
the normalisation is adjusted or entirely determined from the $\beta^* =
2500\un{m}$ data presented in this publication. This allows for different or
even completely independent total cross-section determination with respect to
Ref.~\cite{totem-13tev-90m}.

\subsection{Coulomb-Nuclear Interference}
\label{sec:rho cni}

A detailed overview of different CNI descriptions was given in
Ref.~\cite{totem-8tev-1km}, Section 6. Here we briefly summarise the choices
used for the presented analysis.

The Coulomb amplitude can be derived from QED. In the one-photon approximation
it yields the cross-section
\begin{equation}
\label{eq:coul cs}
	{\d\sigma^{\rm C}\over \d t} = {4\pi\alpha^2\over t^2}\,{\mathcal{F}}^4\ ,
\end{equation}
where $\alpha$ is the fine-structure constant and $\mathcal{F}$ represents an
experimentally determined form factor. Several form factor determinations have
been considered (by Puckett et al., Arrington et al.~and Borkowski et al., see
summary in \cite{elegent}) and no difference in results has been observed.

Motivated by the observed differential cross-section, at low $|t|$ the modulus
of the nuclear amplitude is parametrised as
\begin{equation}
\label{eq:nuc mod}
\left | {\cal A}^{\rm N}(t) \right | = \sqrt{s\over\pi} {p\over \hbar c} \sqrt{a} \exp\left( {1\over 2} \sum\limits_{n = 1}^{N_b} b_n\, t^n \right)\ .
\end{equation}
The $b_1$ parameter is responsible for the leading exponential decrease, the
other $b_n$ parameters can describe small deviations from the leading
behaviour. Since the calculation of CNI may, in principle, involve integrations
(e.g.~Eq.~(\ref{eq:int kl})), it is necessary to extend the nuclear amplitude
meaningfully to higher $|t|$ values, too. In that region, we fix the amplitude
to a function that describes well the dip-bump structure observed in the data.
In order to avoid numerical problems, the intermediate $|t|$ region is modelled
with a continuous and smooth interpolation between the low and high-$|t|$
parts. It has been checked that altering the high-$|t|$ part within reasonable
limits has negligible impact on the results.

Several parametrisations have been considered for the phase of the nuclear
amplitude. Since one of the main goals of this analysis is to compare the newly
obtained $\rho$ value with those at lower energies, we have focused on
parametrisations similar to past analyses. Consequently we have considered
phases with slow variation at low $|t|$: constant, Bailly and standard from
Ref.~\cite{totem-8tev-1km}. No dependence of the results on this choice was
observed and therefore only the constant phase
\begin{equation}
\label{eq:nuc phase con}
\arg {\cal A}^{\rm N}(t) = {\pi\over 2} - \arctan\rho = \hbox{const} \ .
\end{equation}
will be retained in what follows. A more complete exploration including phases
leading to a peripheral description of elastic scattering is planned for a
forthcoming TOTEM publication.

We have used the most general interference formula available in the literature
-- the ``KL'' formula \cite{kl94}:
\begin{equation}
\label{eq:int kl}
	\begin{aligned}
		{\d\sigma\over \d t}^{\rm C+N} =& {\pi (\hbar c)^2 \over s p^2} \left | {\alpha s\over t} {\cal F}^2
			+ {\cal A}^{\rm N}\, \Big[1 - \I\alpha G(t)\Big] \right |^2\ ,\cr
		G(t) =& \int\limits_{-4p^2}^0 \d t'\, \log {t'\over t} {\d\phantom{t'}\over \d t'} {\cal F}^2(t')
			  - \int\limits_{-4p^2}^0 \d t' \left( {{\cal A}^{\rm N}(t') \over {\cal A}^{\rm N}(t)} - 1 \right) { I(t, t')\over 2\pi }\ , \cr
		I(t, t') =& \int_0^{2\pi} \d\phi\ {{\cal F}^2(t'')\over t''}\ ,\cr
		t'' =& t + t' + 2\sqrt{t\, t'} \cos\phi\ ,\cr
	\end{aligned}
\end{equation}
which is numerically almost identical to the formula by Cahn \cite{cahn82} as
shown in Ref.~\cite{totem-8tev-1km}. The CNI effects were calculated by the
computer code from Ref.~\cite{elegent}.

\subsection{Data fits with fixed normalisation}
\label{sec:rho anal}

The fits of the data from Table~\ref{tab:data} have been carried out with the
standard least-squares method, minimising
\begin{equation}
\label{eq:chi sq A}
	\chi^2 = \Delta^\T \mat V^{-1} \Delta\ ,\quad
	\Delta_i = \left.{\d\sigma\over \d t}\right|_{{\rm bin}\ i} - {\d\sigma^{\rm C+N}\over\d t}\left(t^{\rm rep}_{{\rm bin}\ i}\right)\ ,\quad
	\mat V = \mat V_{\rm stat} + \mat V_{\rm syst}\ ,\
\end{equation}
where $\Delta$ is a vector of differences between the differential
cross-section data and a fit function $\d\sigma^{\rm C+N}/\d t$ evaluated at
the representative point $t^{\rm rep}$ of each bin~\cite{lafferty94}. The
minimisation is repeated several times, and the representative points are
updated between iterations. The covariance matrix $\mat V$ has two components.
The diagonal of $\mat V_{\rm stat}$ contains the statistical uncertainty
squared from Table~\ref{tab:data}, $\mat V_{\rm syst}$ includes all systematic
uncertainty contributions except the normalisation, see Eq.~(\ref{eq:covar
mat}). For improved fit stability, the normalisation uncertainty is not
included in the $\chi^2$ definition. In order to propagate this uncertainty to
the fit results, the fit is repeated with the normalisation adjusted by
$+5.5\un{\%}$ and $-5.5\un{\%}$. For each fit parameter the mean deviation from
the fit result with no normalisation adjustment is taken as the effect of
normalisation uncertainty, which is then added quadratically to the uncertainty
reported by the fit with no bias.

The complete fit procedure has been validated with a Monte-Carlo study
confirming that it has negligible bias. It also indicates the composition of
the fit parameter uncertainties. For example, for a fit with $N_b = 1$ using
data in the ``coarse binning'' up to $|t| = 0.07\un{GeV^2}$, the $\rho$
uncertainty due to the statistical uncertainties is about $0.004$, due to the
systematic uncertainties is about $0.003$ and due to the normalisation
uncertainty is about $0.009$.

The fits have been found to have negligible dependence on the binning used (see
Section~\ref{sec:binning}), the choice of electromagnetic form factor (see text
below Eq.~(\ref{eq:coul cs})), the high-$|t|$ nuclear amplitude (see text below
Eq.~(\ref{eq:nuc mod})), the choice of the nuclear amplitude phase (see text
above Eq.~(\ref{eq:nuc phase con})), the number of fit iterations and the
choice of start parameter values for the $\chi^2$ minimisation.

Since the extracted value of $\rho$ may depend on the assumed fit
parametrisation etc., an exploration with various fit configurations has been
performed: several degrees of the hadronic modulus polynomial, $N_b = 1, 2, 3$,
and different sub-samples of the data, constraining them by a maximal value of
$|t|$, $|t|_{\rm max}$. For the latter, two values have been chosen. $|t|_{\rm
max} = 0.15\un{GeV^2}$ corresponds to the largest interval before the
differential cross-section accelerates its decrease towards the dip. It is the
largest interval where application of parametrisation from Eq.~(\ref{eq:nuc
mod}) is sensible. The other choice, $|t|_{\rm max} = 0.07\un{GeV^2}$, reflects
an interval where purely-exponential ($N_b = 1$) nuclear amplitude is expected
to provide a good fit. A summary of the fit results is shown in
Table~\ref{tab:rho ref fits}. The fit with $N_b = 1$ on the larger $|t|$ range
has bad quality, thus the $\rho$ value is not displayed. This shows that the
data are not compatible with a pure exponential, similarly to the previous
observation at $\sqrt s = 8\un{TeV}$ \cite{totem-8tev-90m,totem-8tev-1km}.
Except for this case, all other fit configurations yield good quality and
$\rho$ values constrained to a narrow range.

\begin{table}
\caption{%
Summary of results for various fit configurations (medium binning).
}%
\vskip-5mm
\label{tab:rho ref fits}
\begin{center}
\setlength{\tabcolsep}{5pt}
\begin{tabular}{ccccccccc}
\hline
      & \hbox to10pt{} &\multispan3\hss $|t|_{\rm max} = 0.07\un{GeV^2}$\hss & \hbox to10pt{} & \multispan3\hss $|t|_{\rm max} = 0.15\un{GeV^2}$\hss\cr
$N_b$ && $\chi^2/\hbox{ndf}$ & $\rho$ & $\sigma_{\rm tot}\ung{mb}$ && $\chi^2/\hbox{ndf}$ & $\rho$ & $\sigma_{\rm tot}\ung{mb}$\cr
\hline
\vrule width0pt height10pt
1     && $0.9$ & $0.09\pm0.01$ & $ 111.8 \pm 3.1$  &&     $2.1$ & -              & - \cr
2     && $0.9$ & $0.10\pm0.01$ & $ 111.9 \pm 3.1$  &&     $1.0$ & $0.09\pm0.01$  & $ 111.9 \pm 3.1$\cr
3     && $0.9$ & $0.09\pm0.01$ & $ 111.9 \pm 3.0$  &&     $0.9$ & $0.10\pm0.01$  & $ 112.1 \pm 3.1$\cr
\hline
\end{tabular}
\end{center}
\end{table}

\begin{figure}
\vskip-5mm
\begin{center}
\includegraphics{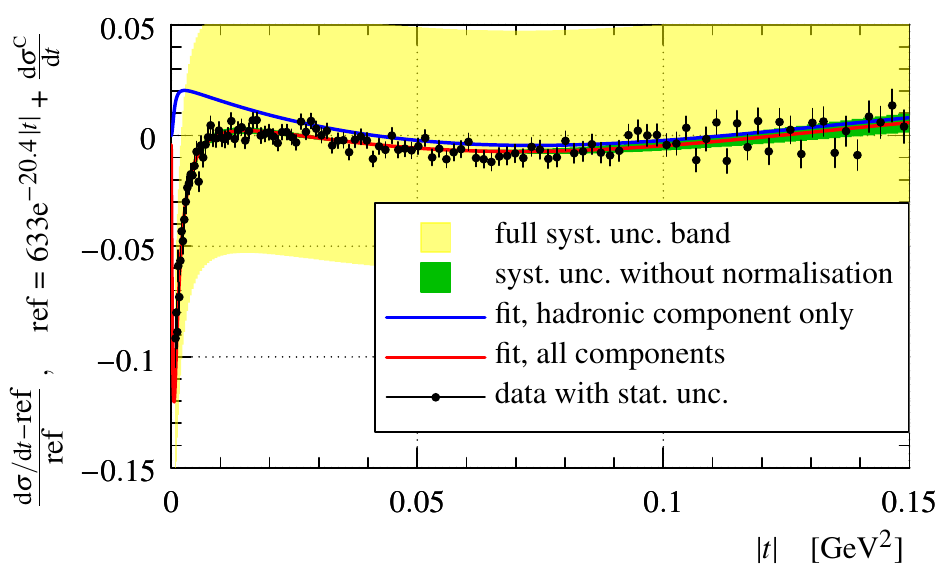}
\vskip-2mm
\caption{%
Details of fit with $N_b = 3$ and $|t|_{\rm max} = 0.15\un{GeV^2}$. The fit parameters read: $a = (648\pm 34)\un{mb/GeV^2}$, $b_1 = (10.64 \pm 0.08)\un{GeV^{-2}}$, $b_2 = (4.1 \pm 1.1)\un{GeV^{-4}}$, $b_3 = (10.3 \pm 4.9)\un{GeV^{-6}}$ and $\rho = 0.10 \pm 0.01$.
}
\label{fig:fit exp3 0.15}
\end{center}
\end{figure}

\begin{figure}
\vskip-5mm
\begin{center}
\includegraphics{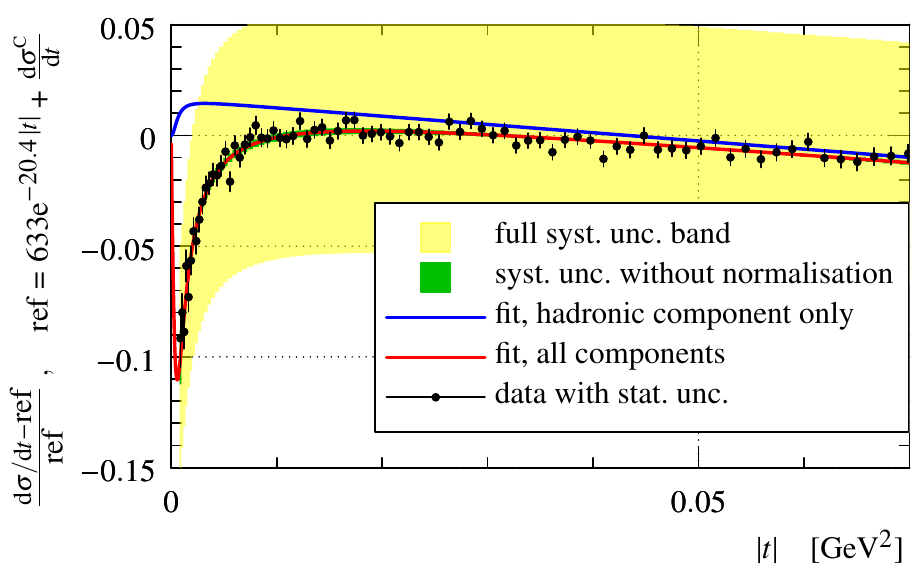}
\vskip-2mm
\caption{%
Details of fit with $N_b = 1$ and $|t|_{\rm max} = 0.07\un{GeV^2}$. The fit parameters read: $a = (643\pm 35)\un{mb/GeV^2}$, $b_1 = (10.39 \pm 0.03)\un{GeV^{-2}}$ and $\rho = 0.09 \pm 0.01$.
}
\label{fig:fit exp1 0.07}
\end{center}
\end{figure}

The extreme cases in Table~\ref{tab:rho ref fits}, combination $N_b=1$ with
$|t|_{\rm max} = 0.07\un{GeV^2}$ and $N_b=3$ with $|t|_{\rm max} =
0.15\un{GeV^2}$ have important meanings. In the latter, the largest possible
sample is used and maximum flexibility is given to the fit. In that sense, this
fit corresponds to the best $\rho$ determination considered. Also, in this case
the fit data include many points where the CNI effects are limited.
Consequently, the fit can ``learn'' the trend of the nuclear component and
``impose it'' in the region of strong CNI effects. Conversely, the fit
configuration $N_b=1$ with $|t|_{\rm max} = 0.07\un{GeV^2}$ relies uniquely on
data with sizeable CNI effects. This complementarity explains why these two
cases give the extreme values of $\rho$ in Table~\ref{tab:rho ref fits}. Fit
details for these two configurations are shown in Figures~\ref{fig:fit exp3
0.15} and \ref{fig:fit exp1 0.07}.

The fit configuration $N_b=1$ with $|t|_{\rm max} = 0.07\un{GeV^2}$ has another
important meaning. Considering the shrinkage of the ``forward-cone'', this
$|t|$ range is similar to the one used in the UA4/2 analysis \cite{ua4-rho}.
This fact may suggest why UA4/2 could not observe deviations of the
differential cross-section from pure exponential: the $|t|$ range was too
narrow, as it would be for the present data, had the acceptance stopped at $|t|
= 0.07\un{GeV^2}$, see Figure~\ref{fig:fit exp1 0.07}. Beyond the $|t|$ range,
this fit combination shares more similarities with the UA4/2 fit (and in
general with many other past experiments): purely exponential fit and
assumption of constant hadronic phase. Moreover, as shown in
Ref.~\cite{totem-8tev-1km}, the ``KL'' interference formula \cite{kl94} used in
this report gives for this fit configuration very similar $\rho$ results as the
``SWY'' interference formula \cite{wy68} used in many past data analyses. From
this point of view this fit combination corresponds to the most fair comparison
to previous $\rho$ determinations and their extrapolations, as e.g.~in Figure
\ref{fig:rho_vs_s}. It is worth noting that this fit configuration yields a
$\rho$ value incompatible at the level of about $4.7\un{\sigma}$ with the
preferred COMPETE model (blue curve in the figure).

\begin{figure*}
\vskip-5mm
\begin{center}
\includegraphics{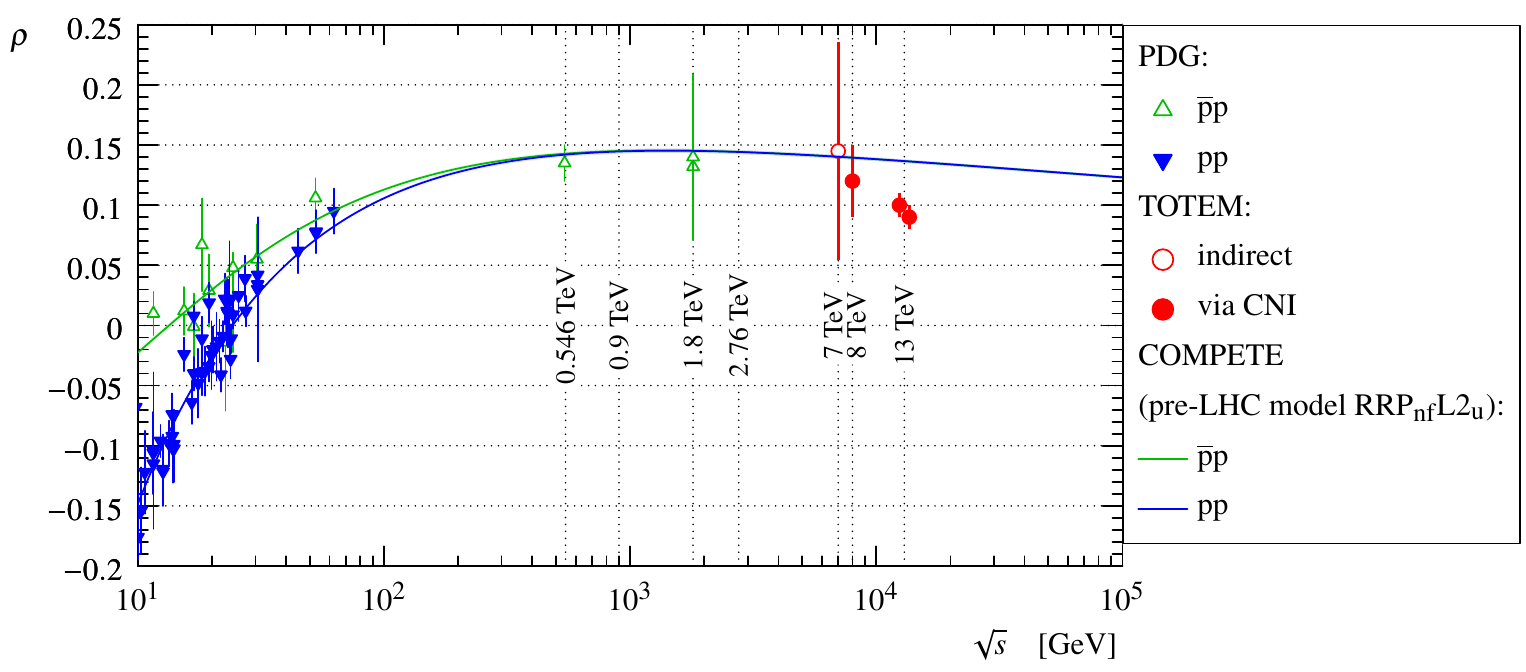}
\caption{%
	Dependence of the $\rho$ parameter on energy. The $\rm pp$ (blue) and
	$\rm p\bar p$ (green) data are taken from PDG \cite{pdg-2010}. TOTEM
	measurements are marked in red. The two points at $13\un{TeV}$
	correspond to the two selected fit cases discussed in text: the
	lhs.~point to the combination $N_b = 3$ and $|t|_{\rm max} =
	0.15\un{GeV^2}$ while the rhs.~point to $N_b = 1$ and $|t|_{\rm max} =
	0.07\un{GeV^2}$.
	}
\label{fig:rho_vs_s}
\end{center}
\end{figure*}

Further tests were performed in order to probe the stability of the $\rho$
extraction. Since at higher $|t|$ values the effects of CNI are limited, one
may conceive a two-step fit: first, use only the higher $|t|$ data to determine
the parameters of the hadronic modulus, cf.~Eq.~(\ref{eq:nuc mod}), and second,
optimise only $\rho$ with all the data but the hadronic modulus fixed from the
first step. Figure~\ref{fig:fit exp3 0.15} indicates that for the first step
one needs to include points down to about $|t| = 0.04\un{GeV^2}$ in order to
describe correctly the concavity of the data. Performing the two-step fit with
$N_b=3$ and with ansatz $\rho = 0.10$ (or $0.14$) yields, at the end, $\rho =
0.103$ (or $0.116$). Although there is a non-zero $\rho$ difference (CNI
effects cannot be fully neglected at higher $|t|$), these results demonstrate
the pull of the data towards $\rho \approx 0.10$. A logical counterpart of the
procedure just described would be to give the higher-$|t|$ data less weight. In
its extreme, where the higher-$|t|$ data are not used at all, this has already
been covered by fits with $|t|_{\rm max} = 0.07\un{GeV^2}$ discussed above,
also showing the preference for lower $\rho$ values.

\begin{figure*}
\vskip-5mm
\begin{center}
\includegraphics{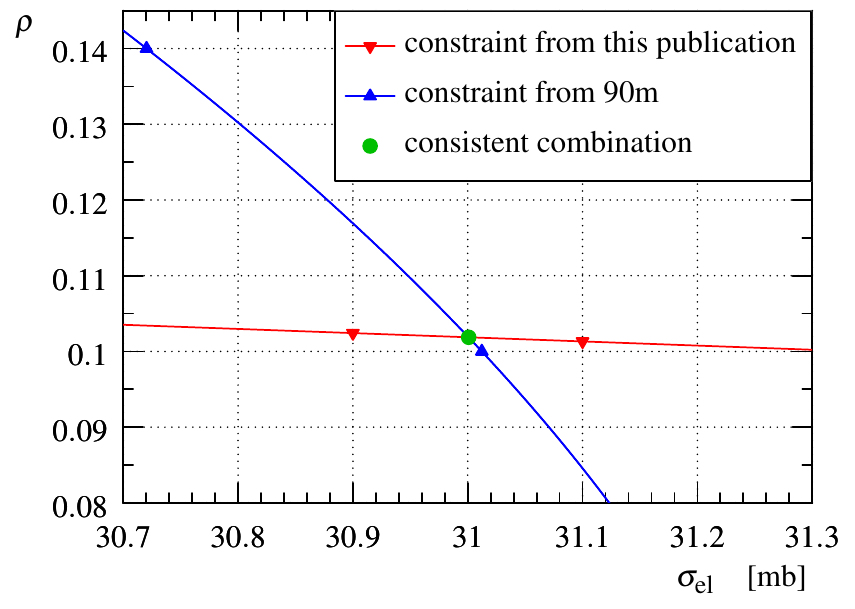}
\vskip-3mm
\caption{%
	Constraints to the relation between $\rho$ and $\sigma_{\rm el}$
	derived from this data set (red line) and from
	Ref.~\cite{totem-13tev-90m} (blue line). The solution consistent with
	both constraints is marked with a green dot.
	}
\label{fig:si_el rho sol}
\end{center}
\vskip-2mm
\end{figure*}

Figure \ref{fig:si_el rho sol} illustrates a small correction due to a
conceptual improvement in combining the data from this publication and from
Ref.~\cite{totem-13tev-90m}. The latter assumes certain values of $\rho$ in
order to evaluate cross-section estimates which are in turn used in this
analysis (see Section~\ref{sec:normalisation}) to estimate $\rho$. This
circular dependence can be resolved by considering simultaneously the $\rho$
dependence of $\sigma_{\rm el}$ in Ref.~\cite{totem-13tev-90m} (blue line) and
the $\sigma_{\rm el}$ dependence of $\rho$ determined in this analysis (red
line). The latter is done as linear interpolation of $\rho$ values extracted
assuming $\sigma_{\rm el} = 30.9$ and $31.1\un{mb}$. The linear dependence is
confirmed with Monte-Carlo studies. The solution consistent with both datasets
(green dot) brings negligible correction to $\rho$ and $-0.03\un{\%}$
correction to the value of $\sigma_{\rm el}$ published in
Ref.~\cite{totem-13tev-90m} for $\rho=0.10$.

For each of the fits presented above, the total cross-section can be derived via the optical theorem:
\begin{equation}
\label{eq:si tot}
\sigma_{\rm tot}^2 = {16\pi\, (\hbar c)^2\over 1 + \rho^2}\, a\ ,
\end{equation}
the results are listed in Table~\ref{tab:rho ref fits}.

\subsection{Data fits with variable normalisation}
\label{sec:rho anal norm var}

Beyond the determination of the $\rho$ parameter, the very low $|t|$ data offer
a normalisation method, too. Suppose that the nuclear amplitude in
Eq.~(\ref{eq:int kl}) were negligible, then the normalisation of the
differential cross-section could be performed with respect to the Coulomb
amplitude, known from QED. While such an extreme situation does not occur
within the available dataset, Table~\ref{tab:data}, the lowest $|t|$ points
receive large contribution from the Coulomb amplitude and can thus be used for
normalisation adjustment or determination. In practice, we extend the fit
function in Eq.~(\ref{eq:int kl}) with parameter $\eta$
\begin{equation}
\label{eq:fit fcn eta}
{\d\sigma^{C+N}\over \d t} = \eta\ {\pi (\hbar c)^2 \over s p^2} \left | {\alpha s\over t} {\cal F}^2
			+ {\cal A}^{\rm N}\, \Big[1 - \I\alpha G(t)\Big] \right |^2\ ,
\end{equation}
which represents normalisation adjustments with respect the $\beta^* =
90\un{m}$ result \cite{totem-13tev-90m} (corresponding to $\eta = 1$).

In turn, the normalisation can be determined from the $\beta^* = 90\un{m}$ data
(Ref.~\cite{totem-13tev-90m} and Section~\ref{sec:normalisation}), from the
$\beta^* = 2500\un{m}$ data (this publication) or their combination. This is
formalised in the following three approaches.
\begin{itemize}
\item approach 1: normalisation from $90\un{m}$ data, results presented in the previous section (in particular Table~\ref{tab:rho ref fits}),
\item approach 2: normalisation estimated with $2500\un{m}$ data under the constraint (mean, RMS) from the $90\un{m}$ data,
\item approach 3: normalisation estimated only from $2500\un{m}$ data.
\end{itemize}

Since the Coulomb normalisation is performed at very low $|t|$, the
presentation in this section will focus on fits with $N_b = 1$. Fits with $N_b
= 3$ were tested, too, without significant changes in the results. For the sake
of simplicity, only the medium binning will be used in this section. The
previous section has shown that results do not depend on the choice of binning.

Since the nuclear-amplitude component cannot be neglected even at the lowest
$|t|$ points of the available dataset, Table~\ref{tab:data}, the normalisation
determination must be performed with care. It has been found preferable to make
the fits in sequence of three steps, using dedicated and physics-motivated fit
configurations for each parameter. The parameters of the nuclear amplitude are
determined from a ``golden nuclear $|t|$ range'' where $|t|$ is large enough
for CNI effects to be small while $|t|$ is small enough for the $N_b = 1$
parametrisation to be suitable. For example, analysing Eq.~(\ref{eq:int kl})
one can find that CNI effects modify the nuclear cross-section by less than
$1\un{\%}$ for $|t| \gtrsim 0.007\un{GeV^2}$. This range agrees with what is
empirically found when trying to go as low as possible in $|t|$ with the
nuclear range without finding significant deviations from the exponential with
$N_b = 1$ either due to the destructive interference with the Coulomb
interaction or due to the non-exponentiality of the nuclear amplitude
\cite{totem-8tev-90m}. In the nuclear range, the CNI effects can be ignored
(charging the residual effects on systematics), making the fit independent of
the interference modelling. The normalisation $\eta$, in contrary, is
determined from the lowest $|t|$ points which are the only ones having
sensitivity to the Coulomb-amplitude component. The $\rho$ parameter is derived
from a $|t|$ range where CNI effects are significant, thus including at least
the complement of the nuclear range, $|t| \lesssim 0.007\un{GeV^2}$. Note that
overlapping $|t|$ ranges are used for determination of $\eta$ and $\rho$.

In detail, approach 2 was implemented via the following sequence of fits.
\begin{itemize}
\item Step a (determination of $b_1$): fit over range $0.005 < |t| <
	0.07\un{GeV^2}$, the CNI effects are ignored. The fit gives a p-value
		of $0.75$.
\item Step b (determination of $\eta$): fit over range $|t| <
	0.0015\un{GeV^2}$, with $b_1$ fixed from step a. The overall $\chi^2$
		receives an additional term $(\eta - 1)^2/ \sigma_\eta^2$,
		$\sigma_\eta = 0.055$, which reflects the constraint from the
		$\beta^* = 90\un{m}$ data. The fit gives negligible average
		pull and yields a p-value of $0.11$.
\item Step c (determination of $\rho$ and $a$): fit over range $|t| <
	0.07\un{GeV^2}$, with $b_1$ fixed from step a and $\eta$ fixed from
		step b. The fit gives a p-value of $0.73$.
\end{itemize}

The $\rho$ and total cross-section results are listed in Table~\ref{tab:rho si
tot summary}. $\eta$ was found to be $1.005$ thus deviating by a fraction of
sigma ($\sigma_\eta = 0.055$) from the $\beta^* = 90\un{m}$ normalisation.

Approach 3 was implemented via the following sequence of fits.
\begin{itemize}
\item Step a (determination of $\eta a^2$ and $b_1$): fit over range $0.0071 <
	|t| < 0.026\un{GeV^2}$. The CNI effects are ignored, therefore the fit
		is only sensitive to the product $\eta a^2$,
		cf.~Eqs.~(\ref{eq:fit fcn eta}) and (\ref{eq:nuc mod}). The fit
		yields a p-value of $0.91$.
\item Step b (determination of $\eta$): fit over range $|t| <
	0.0023\un{GeV^2}$, with $b_1$ and product $\eta a^2$ fixed from step a.
		Since $\eta$ is determined and the product $\eta a^2$ is fixed,
		$a$ is also determined in this step. The fit gives negligible
		average pull and yields a p-value of $0.14$.
\item Step c (determination of $\rho$): fit over range $|t| <
	0.0071\un{GeV^2}$, with $b_1$ fixed from step a and $\eta$ and $a$
		fixed from step b. The fit yields a p-value of $0.23$.
\end{itemize}

The $\rho$ and total cross-section results are listed in Table~\ref{tab:rho si
tot summary}. $\eta$ was found to be $1.020$ thus deviating by less than half a
sigma ($\sigma_\eta$) from the $\beta^* = 90\un{m}$ normalisation. 

As a test we tried approach 3 implementation with a single fit over $|t| <
0.05\un{GeV^2}$, where all parameters ($\eta$, $a$, $b_1$ and $\rho$) are free
and initialised to the values obtained in the previous paragraph. As
anticipated above, such fit might have encountered problems due to non-optimal
parameter sensitivities on the available $|t|$ range, however, the results
listed in Table~\ref{tab:rho si tot summary} are reasonable. $\eta$ was found
to be $1.005$ thus deviating by less than a sigma ($\sigma_\eta$) from the
$\beta^* = 90\un{m}$ normalisation. The fit quality is good: p-value of $0.70$,
see also the illustration in Figure~\ref{fig:approach 3 single fit}.

\begin{figure*}
\vskip-5mm
\begin{center}
\includegraphics{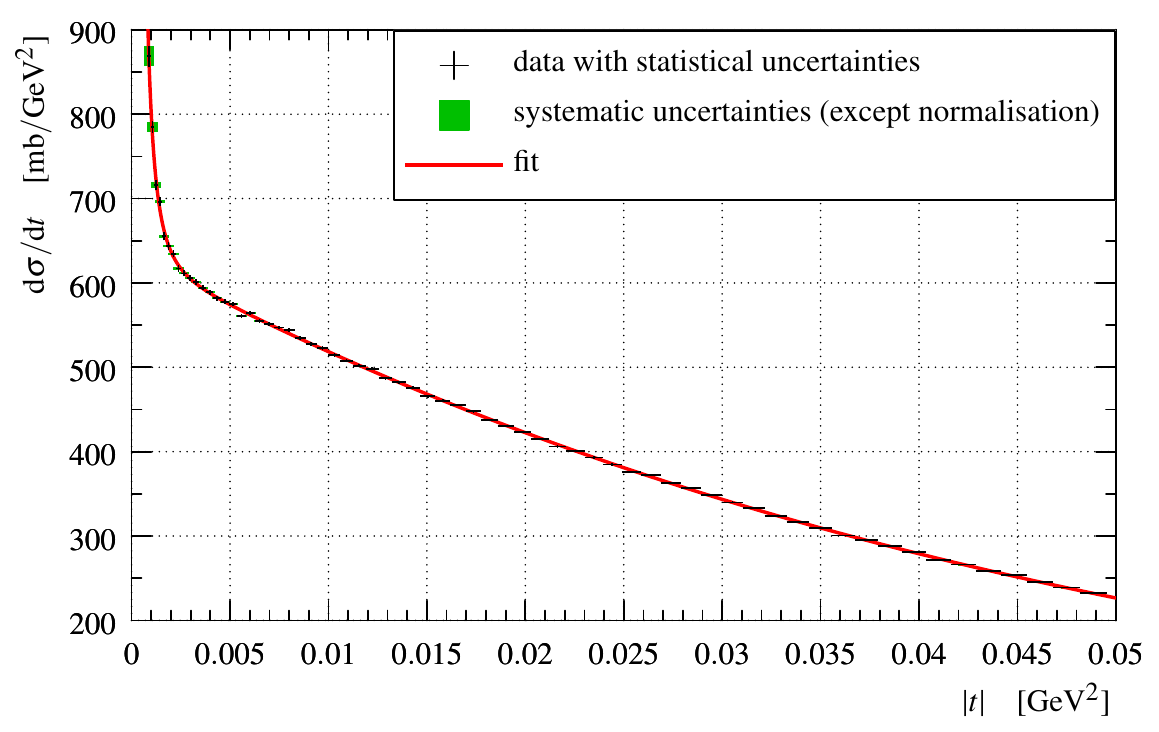}
\vskip-3mm
\caption{
	Illustration of approach 3, single fit. The data come from
	Table~\ref{tab:data}, the normalisation uncertainty is not shown as it
	is not relevant for this fit.
}
\label{fig:approach 3 single fit}
\end{center}
\vskip-3mm
\end{figure*}

\begin{table}
\caption{%
Summary of $\rho$ and total cross-section results.
}%
\vskip-5mm
\label{tab:rho si tot summary}
\begin{center}
\setlength{\tabcolsep}{5pt}
\begin{tabular}{cccc}
\hline
data & method														& $\rho$				& $\sigma_{\rm tot}\ung{mb}$ \cr
\hline
$\beta^* = 90\un{m}$			& Ref.~\cite{totem-13tev-90m}		& -						& $110.6 \pm 3.4$		\cr
\hline
$\beta^* = 2500\un{m}$			& \hbox{approach 1}					& $0.09 \pm 0.01$		& $111.8 \pm 3.2$	\cr
								& \hbox{approach 2}					& $0.09 \pm 0.01$		& $111.3 \pm 3.2$	\cr
								& \hbox{approach 3}					& $0.08(5) \pm 0.01$	& $110.3 \pm 3.5$	\cr
								& \hbox{approach 3 (single fit)}	& $0.10 \pm 0.01$		& $109.3 \pm 3.5$	\cr
\hline
$\beta^* = 90$ and $2500\un{m}$	& Ref.~\cite{totem-13tev-90m} $\oplus$ \hbox{approach 3} &	& $110.5 \pm 2.4$ \cr
\hline
\end{tabular}
\end{center}
\end{table}

The uncertainties for the fits presented above were determined with the
following procedure. The experimentally determined $\d\sigma/\d t$ histogram
was modified by adding randomly generated fluctuations reflecting the
statistical, systematic and normalisation uncertainties (see
Section~\ref{sec:systematics}). This was done $100$ times with different random
seeds. Each of the modified histograms was fitted by the above sequences,
yielding fit parameter samples to determine the parameter fluctuations,
i.e.~uncertainties. Histogram modifications resulting in excessive parameter
deviations from the unmodified fit ($\Delta\rho > 0.05$ or $\Delta\sigma_{\rm
tot} > 10\un{mb}$) were disregarded since such cases would not be accepted in
the analysis. This estimation method gives consistent results with
Section~\ref{sec:rho anal} (for approach 1) and $\chi^2$-based estimate (from
approach 3, single fit). The $\rho$ and $\sigma_{\rm tot}$ uncertainties were
cross-checked and adjusted by varying one of the variables with its uncertainty
at a time for the steps where several variables were determined.

Table~\ref{tab:rho si tot summary} compares $\rho$ and total cross-section
results from Ref.~\cite{totem-13tev-90m} and the approaches described above.
All the results are consistent within the estimated uncertainties. The top two
rows use the same normalisation, which is a decisive component for the total
cross-section value. The larger $\sigma_{\rm tot}$ obtained in this publication
can be attributed to the methodological difference: the destructive
Coulomb-nuclear interference is explicitly subtracted here. The $\sigma_{\rm
tot}$ determinations from Ref.~\cite{totem-13tev-90m} and approach 3 are
completely independent, both in terms of data and method, and can therefore be
combined for uncertainty reduction. The weighted average yields:
\begin{equation}
\label{eq:si tot comb}
\sigma_{\rm tot} = (110.5 \pm 2.4)\un{mb}\ ,
\end{equation}
which corresponds to $2.2\un{\%}$ relative uncertainty.

Figure~\ref{fig:si tot inel el} compares selected total cross-section
measurements at $\sqrt s = 13\un{TeV}$ with past measurements.

\begin{figure*}
\vskip-5mm
\begin{center}
\includegraphics{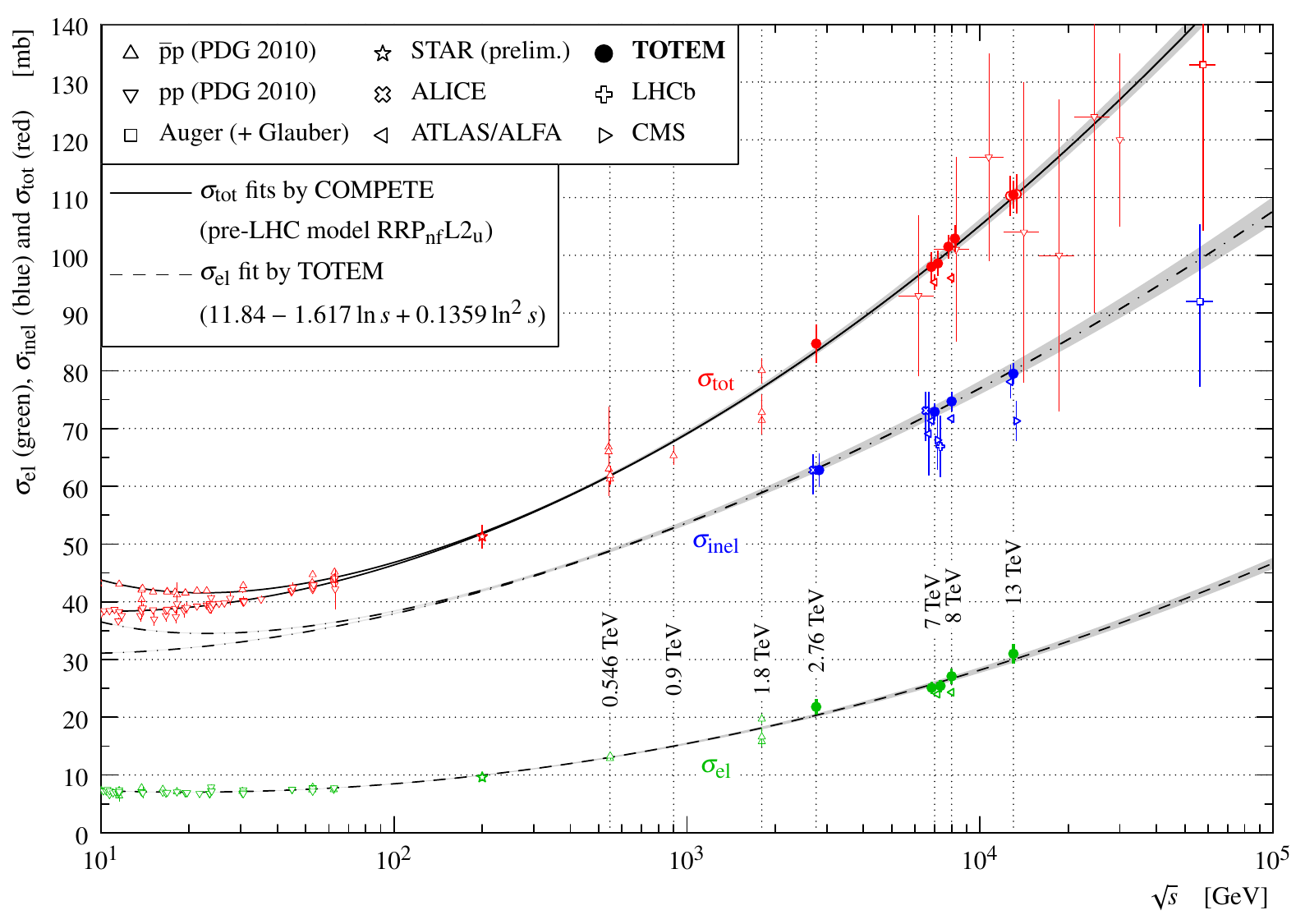}
\vskip-3mm
\caption{%
	Total (red), inelastic (blue) and elastic (green) cross-section as a
	function of energy, $\sqrt s$. The data are taken from
	Ref.~\cite{totem-13tev-90m} (and references therein) and
	Table~\ref{tab:rho si tot summary}. At $\sqrt s = 13\un{TeV}$, three
	total cross-section points are shown: left hollow corresponds to
	approach 3, right hollow to Ref.~\cite{totem-13tev-90m} and central
	filled to the average in Eq.~(\ref{eq:si tot comb}).
	}
\label{fig:si tot inel el}
\end{center}
\vskip-3mm
\end{figure*}

\section{Discussion of Physics Implications}
\label{sec:discussion}

One very comprehensive (and therefore representative) studies of the pre-LHC
data is by the COMPETE collaboration \cite{compete}. In total 256 models, all
without crossing-odd components, were considered to describe $\sigma_{\rm tot}$
and $\rho$ data for various reactions ($\rm pp$, $\rm p\pi$, $\rm pK$, etc.)
and the corresponding particle-antiparticle reactions. Out of these models, 23
were found to give a reasonable description of the data \cite{compete-details}.
Extrapolations from these models are confronted with newer TOTEM measurements
in Figure~\ref{fig:comp bands}, which shows that they are grouped in 3 bands.
Each band is plotted in a different colour and has a different level of
compatibility with the data. As argued above, the $13\un{TeV}$ fit with $N_b=1$
and $|t|_{\rm max} = 0.07\un{GeV^2}$ (rightmost point in the figure)
corresponds to the most fair comparison to past analyses and is therefore used
to evaluate the compatibility with the COMPETE models. The $8\un{TeV}$ $\rho$
point is not included in this calculation since it does not bring any
information due to its large uncertainty. The $\sigma_{\rm tot}$ measurements
can be, to a large extent, regarded as independent: they used data from
different LHC fills at different energies, different beam optics, often
different RPs, often different analysis approaches (fit parametrisation,
treatment of CNI) and often they were analysed by different teams. The only
correlation comes from using common normalisation at a given collision energy.
Consequently, two compatibility evaluations were made: using all $\sigma_{\rm
tot}$ points from Figure~\ref{fig:comp bands} and using their subset with a
single point per energy. These two results thus provide upper and lower bounds
for the actual compatibility level. The observations can be summarised as
follows.

\begin{itemize}[noitemsep,topsep=0pt]
	\item The blue band is compatible (p-value $0.990$ to $0.995$) with the
		$\sigma_{\rm tot}$ data, but incompatible (p-value
		$3\times10^{-6}$) with the $\rho$ point.
	\item The magenta band is incompatible (p-value $1\times10^{-5}$ to
		$5\times10^{-4}$) with the $\sigma_{\rm tot}$ data and
		incompatible (p-value $9\times10^{-3}$) with the $\rho$ point.
	\item The green band is incompatible (p-value $3\times10^{-18}$ to
		$5\times10^{-12}$) with the $\sigma_{\rm tot}$ data, but
		compatible (p-value $0.4$) with the $\rho$ point.
\end{itemize}
In summary, none of the COMPETE models is compatible with the ensemble of
TOTEM's $\sigma_{\rm tot}$ and $\rho$ measurements.

\begin{figure*}
\vskip-5mm
\begin{center}
\includegraphics{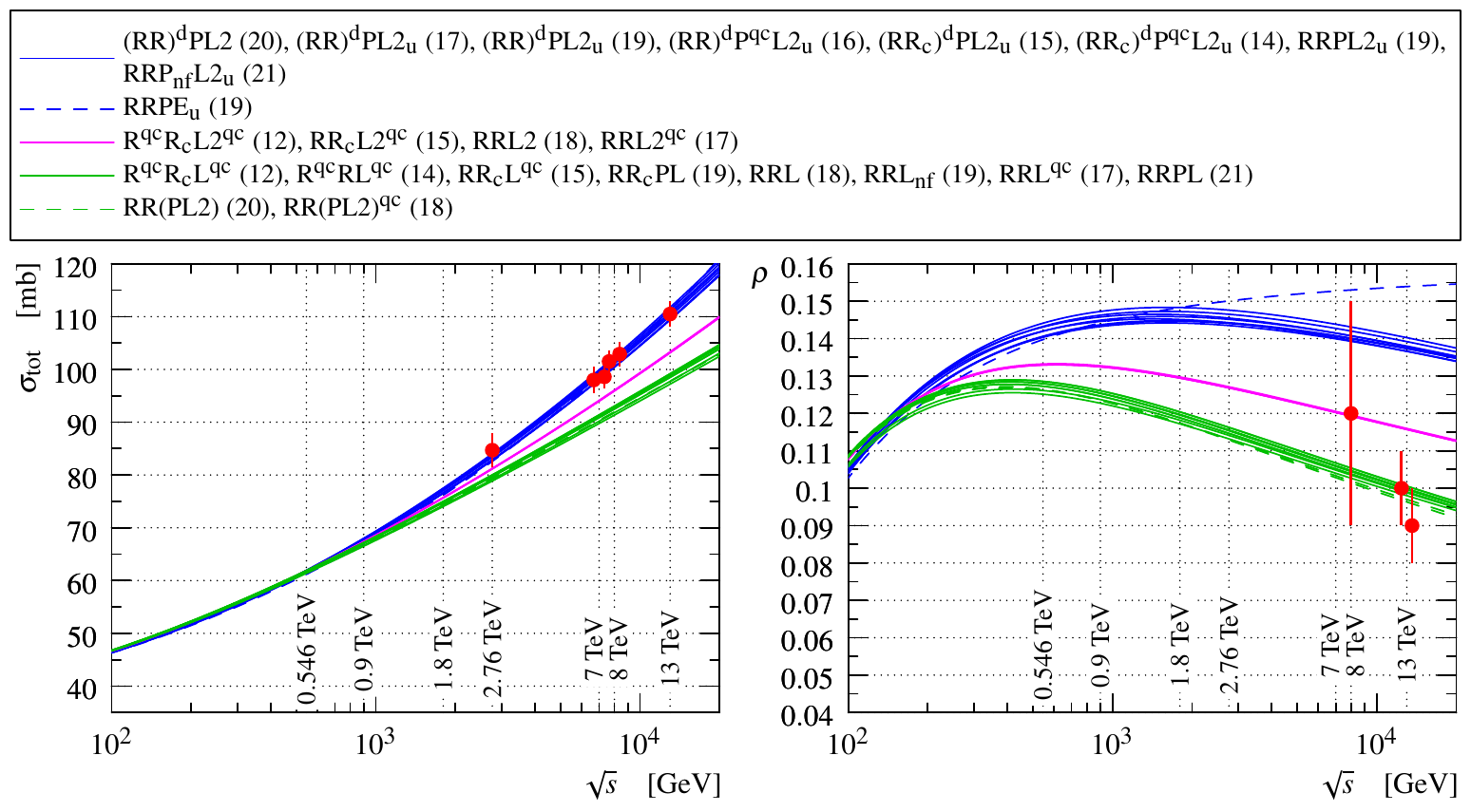}
\caption{%
	Predictions of COMPETE models \cite{compete-details} for $\rm pp$
	interactions. Each model is represented by one line (see legend). The
	red points represent the reference TOTEM measurements. The $\sigma_{\rm
	tot}$ point at $13\un{TeV}$ corresponds to the weighted average in
	Eq.~(\ref{eq:si tot comb}). The two $\rho$ points at $13\un{TeV}$
	correspond to the two cases discussed in Section~\ref{sec:rho anal}:
	the left point to the fit with $N_b=3$ and $|t|_{\rm max} =
	0.15\un{GeV^2}$, the right point to $N_b=1$ and $|t|_{\rm max} =
	0.07\un{GeV^2}$.
	}
\label{fig:comp bands}
\end{center}
\end{figure*}

Another, even less model-dependent, relation between $\sigma_{\rm tot}$ and
$\rho$ can be obtained from dispersion relations
\cite{dremin-dispersion,barone-predazzi}. If only the crossing-even component
of the amplitude is considered, it can be shown that $\rho$ is proportional to
the rate of growth of $\sigma_{\rm tot}$ with energy. Therefore, the low value
of $\rho$ determined in Section~\ref{sec:rho} indicates that either the total
cross-section growth should slow down at higher energies or that there is a
need for an odd-signature object being exchanged by the protons. While at lower
energies such contributions may naturally come from secondary Reggeons, their
contribution is generally considered negligible at LHC energies due to their
Regge trajectory intercept lower than unity.

A variety of odd-signature exchanges relevant at high energies have been
discussed in literature, within different frameworks and under different names,
see e.g.~the reviews \cite{braun,ewerz}. The ``Odderon'' was introduced within
the axiomatic theory \cite{nicolescu-1973,nicolescu-1992,nicolescu-2007} as an
amplitude contribution responsible for $\rm p\bar p$ vs.~$\rm pp$ differences
in the total cross-section as well as in the differential cross-section,
particularly in the dip region. Crossing-odd trajectories were also studied
within the framework of Regge theory as a counterpart of the crossing-even
Pomeron. It has also been shown that an such object must exist in QCD, as a
colourless bound state of three gluons with quantum numbers $J^{PC} = 1^{--}$
(see e.g.~\cite{bartels-2000}). The binding strength among the 3 gluons is
greater than the strength of their interaction with other particles. There is
also evidence for such a state in QCD lattice calculations, known under the
name ``vector glueball'' (see e.g.~\cite{morningstar-1999}). Such a state, on
one hand, can be exchanged in the $t$-channel and contribute, e.g., to the
elastic-scattering amplitude. On the other hand it can be created in the
$s$-channel and thus be observed in spectroscopic studies. Non-perturbative QCD
studies based on the AdS/CFT correspondence show that the Odderon emerges on
equally firm footing as the Pomeron \cite{brower-2009}.

There are multiple ways how an odd-signature exchange component may manifest
itself in observable data. Focussing on elastic scattering at the LHC
(unpolarised beams), there are 3 regions often argued to be sensitive. In
general, the effects of an odd-signature exchange (3 bound gluons) are expected
to be much smaller than those of even-signature exchanges (2 bound gluons).
Consequently, the sensitive regions are those where the contributions from
2-bound-gluon exchanges cancel or are small. At very low $|t|$ the
2-bound-gluon amplitude is expected to be almost purely imaginary, while a
3-gluon exchange would make contributions to the real part and therefore $\rho$
is a very sensitive parameter. Another such example is the dip, often described
as the imaginary part of the amplitude crossing zero, thus ceding the dominance
to the real part to which a 3-gluon exchange may contribute. In agreement with
such predictions, the observed dips in $\rm p\bar p$ scattering are shallower
than those in $\rm pp$. At $\sqrt s = 53\un{GeV}$, there are data showing a
very significant difference between the $\rm pp$ and $\rm p\bar p$ dip
\cite{breakstone-85}. The interpretation of this difference is, however,
complicated due to non-negligible contribution from secondary Reggeons. These
are not expected to give sizeable effects at the Tevatron energies, which thus
gives weight to the D0 observation of a very shallow dip in $\rm p\bar p$
elastic scattering \cite{d0-elastic} compared to the very pronounced dip
measured by TOTEM at $7\un{TeV}$ \cite{totem-7tev-first}. The $\rm pp$ vs.~$\rm
p\bar p$ dip difference is also predicted to be energy-dependent which presents
another experimental observable (see e.g.~\cite{ster-2015}). Sometimes the
high-$|t|$ region is also argued to be sensitive to 3-gluon exchanges since the
contribution from 2-gluon exchanges is rapidly dropping. However, preliminary
high-$|t|$ TOTEM data at $13\un{TeV}$ indicate that this region is dominated by
a perturbative-QCD amplitude, e.g.~\cite{Donnachie:1979yu}.

\begin{figure*}
\vskip-5mm
\begin{center}
\includegraphics{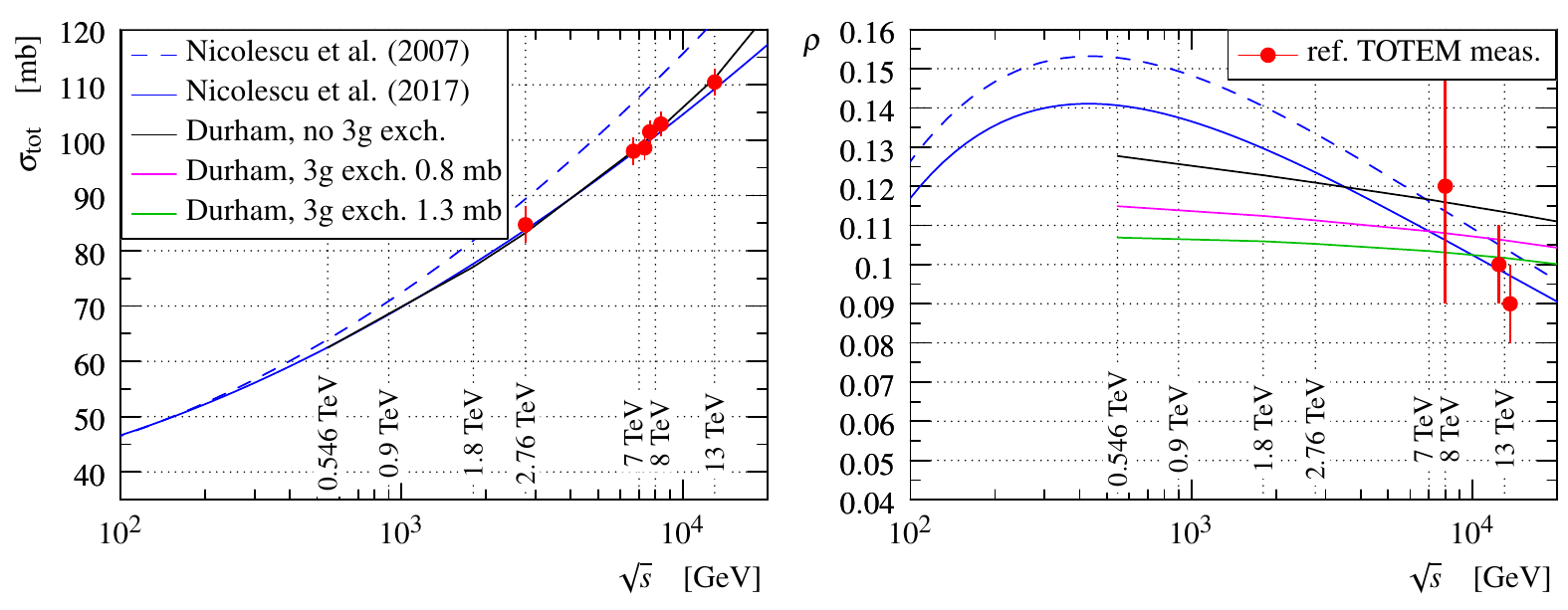}
\caption{%
	Predictions of the model by Nicolescu et al.~(dashed blue curve from
	\cite{nicolescu-2007}, solid blue curve from \cite{nicolescu-2017}) and
	the Durham model \cite{durham-2018} (including crossing-odd
	contribution from \cite{levin-1990}) compared to the reference TOTEM
	measurements (red dots). The $\sigma_{\rm tot}$ point at $13\un{TeV}$
	corresponds to the weighted average in Eq.~(\ref{eq:si tot comb}). The
	two $\rho$ points at $13\un{TeV}$ correspond to the two cases discussed
	in Section~\ref{sec:rho anal}: the left point to the fit with $N_b=3$
	and $|t|_{\rm max} = 0.15\un{GeV^2}$, the right point to $N_b=1$ and
	$|t|_{\rm max} = 0.07\un{GeV^2}$. For the Durham model the black curve
	corresponds to the prediction without a colourless 3-gluon $t$-channel
	exchange. The magenta and green curves refer to the $\rm pp$
	predictions including a 3-gluon exchange with proton coupling
	equivalent to $0.8$ and $1.3\un{mb}$, respectively.
	}
\label{fig:match models}
\end{center}
\end{figure*}

Figure~\ref{fig:match models} compares the TOTEM data with two compatible
models: by Nicolescu et al.~\cite{nicolescu-2017} and the extended Durham model
\cite{durham-2018} (original model \cite{durham-2014} plus crossing-odd
contribution from \cite{levin-1990}). The 2007 version of the Nicolescu model
(dashed blue) is based only on pre-LHC data and predicts $\sigma_{\rm tot}$
overestimating the TOTEM measurements -- as argued in
Ref.~\cite{nicolescu-2017} it might be due to the ambiguities in prolonging the
amplitudes in the non-forward region. The 2017 version (solid blue) includes
also LHC measurements up to $13\un{TeV}$ and describes the $\sigma_{\rm tot}$
data well. Both versions yield similar results for $\rho$, with a pronounced
energy dependence. This comes from the fact that the crossing-odd component is
almost negligible at $\sqrt s \approx 500\un{GeV}$ but very significant at
$13\un{TeV}$. Conversely, in the Durham model the effect is sizeable at $\sqrt
s \approx 500\un{GeV}$ and gently diminishes with energy. The Durham model also
predicts a mild energy dependence of the $\rho$ parameter. Therefore, precise
$\rho$ measurements at $\sqrt s \approx 900\un{GeV}$ and $14\un{TeV}$ would be
valuable for discrimination between these models. For both models, the
inclusion of a crossing-odd exchange component was essential to reach the
agreement between the data and model. In particular, the Durham model without
such a contribution (black line) is not well compatible (p-value $0.02$) with
the (rhs.) $\rho$ point obtained with $N_b=1$ and $|t|_{\rm max} =
0.07\un{GeV^2}$. 

\section{Summary}
\label{sec:summary}

The measurement of elastic differential cross-section disfavours the 
purely-exponential low-$|t|$ behaviour at $\sqrt s = 13\un{TeV}$, similarly to
the previous observation at $8\un{TeV}$. Thanks to the very low-$|t|$ reach, 
the first extraction of the $\rho$ parameter at $\sqrt s = 13\un{TeV}$ was made
by exploiting the Coulomb-nuclear interference. The fit with conditions similar
to past experiments yields $\rho = 0.09 \pm 0.01$, one of the most precise
$\rho$ determinations in history. The fit over the maximum of data points and
with maximum reasonable flexibility of the fit function gives $\rho = 0.10 \pm
0.01$.

Also thanks to the very low $|t|$ reach, it was possible to apply the ``Coulomb 
normalisation'' technique for the first time at the LHC and obtain another
total cross-section measurement $\sigma_{\rm tot} = (110.3 \pm 3.5)\un{mb}$
completely independent from the previous TOTEM measurement at $\sqrt s =
13\un{TeV}$ \cite{totem-13tev-90m} but well compatible with it. Since these two
measurements are independent, it is possible to calculate the weighted average
yielding $\sigma_{\rm tot} = (110.5 \pm 2.4)\un{mb}$.

The updated collection of TOTEM's $\sigma_{\rm tot}$ and $\rho$ data presents a 
stringent test of model descriptions. For an indicative example, none of the
models considered by the COMPETE collaboration is compatible with both
$\sigma_{\rm tot}$ and $\rho$.

For both models found to be consistent with TOTEM's data, the inclusion of a 
3-gluon-state exchange in the $t$-channel was essential for reaching the good
agreement with the data.

If it is demonstrated in future that the crossing-odd exchange component is 
unimportant for elastic scattering, the low $\rho$ value determined in this
publication represents the first experimental evidence for slowing down of the
total cross-section growth at higher energies, leading to a deviation from most
current model expectations.

\section*{Acknowledgements}

This work was supported by the institutions listed on the front page and also
by the Magnus Ehrnrooth foundation (Finland), the Waldemar von Frenckell
foundation (Finland), the Academy of Finland, the Finnish Academy of Science
and Letters (The Vilho, Yrj\"o and Kalle V\"ais\"al\"a Fund), the OTKA NK
101438 and the EFOP-3.6.1-16-2016-00001 grants (Hungary). Individuals have
received support from Nylands nation vid Helsingfors universitet (Finland),
from the M\v SMT \v CR (Czech Republic) and the J\'anos Bolyai Research
Scholarship of the Hungarian Academy of Sciences and the NKP-17-4 New National
Excellence Program of the Hungarian Ministry of Human Capacities.


\bibliographystyle{h-elsevier.bst}
\bibliography{bibliography}

\end{document}